%% file: IPV_Preferences_arXiv.tex
\documentclass[12pt]{article}
\usepackage[a4paper,
            bindingoffset=0.2in,
            left=1.25in,
            right=1.25in,
            top=1.25in,
            bottom=1.25in]{geometry}
\usepackage[english]{babel}
\usepackage[utf8]{inputenc}
\usepackage{amssymb}
\usepackage{amsfonts}
\usepackage{ragged2e}
\usepackage{amsmath}
\usepackage{amsthm}
\usepackage{eurosym}
\usepackage{bm}
\usepackage{booktabs,caption,fixltx2e}
\usepackage{threeparttable}
\usepackage[pdftex]{graphicx}
\usepackage{rotating}
\usepackage{multirow}
\usepackage{array}
\usepackage{caption,subcaption}
\usepackage[capposition=top]{floatrow}
\usepackage{pdfpages}
\usepackage{pdflscape}
\usepackage{setspace}
\usepackage[normalem]{ulem}
\usepackage{float}
\usepackage{afterpage}
\usepackage{pifont}
\usepackage{appendix}
\usepackage[round, semicolon, authoryear, sort]{natbib}
\usepackage{hyperref}
\usepackage{color}
\usepackage{enumitem}
\usepackage{longtable}
\usepackage{setspace}
\usepackage{tabularx}

\usepackage{footnote}
\usepackage{accents}
\usepackage{xcolor}
\usepackage{tikz}
\usetikzlibrary{calc,arrows}
\usepackage{verbatim}
\usetikzlibrary{fadings}
\newtheoremstyle{mystyle}
{0.4in}
{0in}
{\itshape}
{}
{\bfseries}
{:}
{\newline}
{}
\theoremstyle{mystyle}

\bibpunct{(}{)}{;}{a}{,}{,}
\definecolor{darkmidnightblue}{rgb}{0.0, 0.2, 0.4}
\hypersetup{
colorlinks=true, linkcolor=darkmidnightblue, citecolor=darkmidnightblue, filecolor=darkmidnightblue, urlcolor=darkmidnightblue }
\newcommand{\sym}[1]{\rlap{#1}}

\usepackage{siunitx} 
\begin{document}

\title{\vspace{-2cm}\Large Intimate partner violence and women's economic preferences\footnote{ This research project was funded by the British Academy and the Center for the Economic Analysis of Risk (CEAR), Georgia State University internal grant. U.S. Agency for International Development (Cooperative Agreement Number RFA 72DFFP20RFA00006). Ethical approval for Experiment 1 was obtained from Social Sciences and Humanities Interdivisional Research Ethics Committee (IDREC), University of Oxford [ref. R57949/RE001]. Ethical approval for Experiment 2 was obtained from International Food Policy Research Institute ref. IFPRI PHND-22-0631MM on July 09, 2023 and Ethiopian Public Health Association ref. EPHA/OG/724/23 on July 13, 2023. Experiments are pre-registered at AEA and OSF registries: AEARCTR-0006165 and AEARCTR-0014038.}}

\author{%
Dan Anderberg\thanks{Royal Holloway University of London} \and Rachel Cassidy\thanks{Institute for Fiscal Studies \& World Bank Group}  \and Anaya Dam \thanks{%
Utrecht University \& University of Oxford. Corresponding Author. Email: \href{mailto:a.dam@uu.nl}%
{a.dam@uu.nl}} \and Melissa Hidrobo\thanks{International Food Policy Research Institute}%
\and Jessica Leight\thanks{International Food Policy Research Institute} \and Karlijn Morsink \thanks{%
Utrecht University, Wageningen University,  Cornell University \& Center for Economic Analysis of Risk (CEAR), Georgia State University.}}

\date{\small\today}
\maketitle
\vspace{\fill}
\vspace{-1.3cm}
\begin{abstract}
\small
\noindent
One in three women globally experiences intimate partner violence (IPV), yet little is known about how such trauma affects economic decision-making. We provide causal evidence that IPV influences women’s time preferences --- a key parameter in models of savings, investment, and labor supply. We combine two empirical strategies using four distinct datasets. First, in two randomized recall experiments in Ethiopia, we randomly assigned women to recall specific acts of abuse before eliciting their intertemporal choices. Women with IPV experiences prompted to recall IPV display significantly greater impatience than otherwise similar women who are not prompted. Second, we exploit exogenous reductions in IPV generated by two randomized interventions—one involving cash transfers, the other psychotherapy—and use treatment assignment as an instrument for IPV exposure. Women who experience reduced IPV as a result of treatment exhibit more patient time preferences. Together, these results provide consistent, novel causal evidence that exposure to IPV induces individuals to discount the future more heavily. This evidence suggests a psychological channel through which violence can perpetuate economic disadvantage and constrain women’s ability to take actions—such as saving, investing, or exiting abusive relationships—that require planning over time.

\end{abstract}
\vspace{-0.1cm}
\vspace{\fill}
\thispagestyle{empty}

\begin{spacing}{1.4}

\newpage
\setcounter{page}{1}

\section{Introduction}
\label{sec:introduction}

Intimate-partner violence (IPV) remains a pervasive global health and human-rights concern: roughly one in three women worldwide reports having endured emotional, physical, or sexual aggression by a current or former partner \citep{UNWomen:2025}. Beyond its immediate physical harms, IPV is associated with elevated psychological distress, post‑traumatic stress disorder (PTSD), functional impairment, and suicidal-thoughts and attempts \citep{Ellsbergetal:2008, Dillonetal:2013, Orametal:2022}. Yet, surprisingly little is known about how the trauma of IPV shapes the preferences that guide economic decision‑making.\footnote{One exception is a recent paper by \citet{Adamsetal:2024}, who show that women cohabiting with an abusive partner face substantial economic costs in terms of income and employment. Less is, however, known about the mechanisms driving these costs, which could work through direct sabotage or trauma-induced changes to preferences.}

A growing literature links exposure to war, natural disasters, and other traumatic shocks to shifts in fundamental economic preferences -- especially risk preferences and time preferences \citep{Eckeletal:2009, Voorsetal:2012, Cassaretal:2017, Callenetal2014}. Three complementary mechanisms have been proposed to explain why trauma might specifically affect individuals' time discounting. The mental-experiencing hypothesis posits that trauma hampers one's capacity to visualize distant outcomes, thereby biasing choices towards the present \citep{Ashrafetal:2024, kleimetal:2014}. The cognitive-impairment hypothesis highlights trauma-related dysfunction in the pre-frontal cortex that underpins impulse control \citep{Twamleyetal:2009, Aupperleetal:2016, Steinetal:2002, Scottetal:2015}. Finally, the cognitive-scarcity hypothesis argues that living in situations of acute stress and negative affective states consumes finite cognitive resources, leaving less bandwidth for future-oriented decisions \citep{haushoferFehr:2014, Muravenetal:2000, Bartovsetal2021}. Whether -- and to what extent -- these pathways operate in the context of IPV remains an open question.

We provide the first causal evidence that IPV alters women’s inter-temporal choices. Combining two pre‑registered survey experiments with instrumental‑variables analyses of two randomized IPV‑reduction programs (four independent datasets with in total N=3076), 
we isolate exogenous variation in both the salience and the incidence of recent IPV. When women who experienced IPV in the preceding twelve months are randomly prompted to recall specific acts and injuries -- using a well-established and commonly used survey-instrument --  they exhibit significantly higher impatience than equally victimized women who are not prompted to recall those events. Consistently, assignment to IPV‑reduction interventions that lowered the probability of abuse by almost 20\%, increased women’s patience in standard monetary choice tasks that are typically used to elicit time preferences. Together these results indicate that IPV can shift victims’ time preferences -- key parameters that influence saving, investment in human capital, and labor‑market participation -- and suggest a novel psychological channel through which violence perpetuates economic dependence.

Causal identification of the effects of IPV on time preferences poses inherent challenges, as IPV cannot be ethically or experimentally assigned. We therefore draw on two complementary second-best strategies. The first is \textit{randomized recall}, a common approach in psychology and behavioral economics \citep{Manietal2013, Bartovsetal2021, Callenetal2014}, in which individuals are randomly prompted to recall a past experience—in this case, IPV. Unlike correlational designs, this method preserves balance on both exposure and unobserved characteristics. However, concerns remain regarding the interpretation of the recall exercise, whether recall of IPV represents the effect of actual IPV trauma, and generalizes to real-life preferences and behaviour. As a second strategy, we use \textit{instrumental variables} designs that exploit exogenous reductions in IPV from randomized interventions, such as cash transfers or cognitive behavioral therapy \citep{TolEtal:2019, HaushoferRingdalShapiro:2019, Hidroboetal:2016, Heathetal2020,Royetal:2022, HaushoferShapiro2016}. While the assignment of such interventions is random by design, potential violations of the exclusion restriction remain a valid concern.

The two pre-registered randomized recall experiments were conducted in Ethiopia. The first experiment consists of 400 women in urban and semi-urban communities with households relying on factory-work for income. The second experiment consists of 1,630 
 women in rural communities with households relying on agriculture and livestock for income. Reported experiences and acceptability of IPV are high in Ethiopia: 63\% of adults say they believe IPV is justified in at least one of five situations asked about in the Demographic and Health Survey (DHS) \citep{CSACE2016}.\footnote{The question asks respondents whether they think it is justified for a husband to beat their wife in the each of the following five situations. If she: ``\textit{goes out without telling the husband}"; ``\textit{neglects the children};" ``\textit{argues with the husband};" ``\textit{refuses to have sex with the husband};" or ``\textit{burns the food}.''}
We randomly assign respondents to one of two experimental arms. Respondents in the treatment arm are asked to recall episodes of emotional, physical and sexual IPV they may have experienced, the frequency of these events, and any injuries they may have incurred, using a survey instrument based on the well-established DHS and WHO instruments.\footnote{See Table \ref{tab:DHS_instrument} for the items and categorization.} Respondents in our control arm receive a placebo module where they are asked to think about scenarios and questions that aimed to mimic casual ‘chit-chat’ or small-talk, based on extensive piloting. 

We elicit respondents' time preferences right after they complete the treatment or control module, using an incentivised multiple price list over a monetary domain \citep{CollerWilliams:1999, Cohenetal:2020}. The respondent is asked to make choices over two different frames in the future. In the near frame they are asked to choose between receiving 50 Birr in 1 day (3 days) or a varying amount in 14 days (10 days) and in the far frame they are asked to choose between receiving 50 Birr in 14 days or a varying amount in 28 days (21 days).\footnote{50 Ethiopian Birr amounts to approximately 0.5 USD, representing modest but meaningful stakes in this context.} Our primary outcomes of interest are the monetary values at which the respondent is indifferent between money received on the earlier and later date in both frames. We extensively piloted the task and trained the enumerators to ensure comprehension among a low-literacy sample. Respondents demonstrated high levels of task comprehension and consistency in responses. After completion of the priming modules and the incentivized experiments, women assigned to the control arm were also asked the IPV module, such that data on IPV was collected from all participants.\footnote{IPV experiences are balanced across the treatment and control arm.}

We find that women who have experienced IPV events in the past months, who are asked to recall these events, demonstrate more impatient behaviour than women with IPV experiences who are not asked to recall these IPV events. Decomposing the effects by the standard categorization of violence in the literature --- emotional, and physical and sexual violence, we find that the effects are driven by emotional violence. Emotional violence encompasses controlling behaviour and threats of physical or sexual harm, rather than the physical or sexual act itself. Point estimates are similar in the near frame (early date: 1 day (3 days) from now) and far frame (early date: 14 days from now), with the effect being significant in the far frame (p-value $<$ 0.10) and marginally insignificant in the near frame (\textit{p-}value $=$ 0.134). The effect sizes suggest that women who have experienced emotional IPV when asked to recall IPV events are willing to forgo 8-11 Birr to receive the amount of money on the earlier day, as compared to women with similar IPV experiences who are not asked to recall IPV events. This amounts to forgoing 10\% of the average present equivalent in the control group of 88.5 Birr. Finally, for women who have not experienced IPV, being randomly asked to recall IPV experiences does not lead to significant impacts on their time preferences, in the near frame or far frame as well as for different categories of violence. We rule out liquidity constraints, and a preference for not receiving payments in the future -- for example, due to low trust in the experimenters, or fear of payments around a violent partner -- as alternative interpretations of our findings.

For our second empirical strategy to achieve causal identification of the effect of IPV on time preferences -- the instrumental variables approach --  the first dataset we leverage is from a randomized control trial of an unconditional cash transfer intervention
in Kenya \citep{HaushoferShapiro2016} with a sample of 1010 cohabiting women. The intervention led to a significant reduction in IPV –- particularly driven by a reduction in physical and sexual violence. We find
that a reduction of one act of IPV in the past 6 months due to the intervention leads to an increase in patience in the near frame (tomorrow versus 6 months) and far frame (six versus twelve months), which is significant in the near frame (p-value $<$ 0.10) but marginally insignificant in the far frame (p-value $=$ 0.125). A reduction in IPV makes women willing to forgo 9-12 less Kenyan Shillings on the earlier day in order to receive a larger sum of money at the later date, amounting to approximately
20\% of the control group mean.\footnote{50 Kenyan Shillings amounts to approximately 0.5 USD, a modest but meaningful amount in this context.} Effects are robust to using a two-stage least squares or median unbiased limited information maximum likelihood estimator, as well as Anderson-Rubin confidence intervals for weak instruments.

The second field dataset we leverage is from a randomized control trial of a psychotherapy intervention in Ethiopia with a sample of 1666 cohabiting women. 
We find that the intervention led to a significant reduction in IPV – particularly driven by a reduction in emotional violence. We find that a reduction in IPV -- from any IPV to no IPV, or from any emotional IPV to no emotional IPV -- leads to a negative point estimate on time preferences in both the near and the far frame, and the effect is marginally insignificant in the latter  (\textit{p}-value is 0.119 and 0.149, respectively).  A full removal of any IPV and emotional IPV due to the intervention makes women willing to wait for 58-88 Birr more on the later date rather than accepting the 50 Birr (smaller amount) on the earlier date. The effects appear noisily estimated, likely due to a weak instrument. Since the analysis was pre-specified, we take the these results as a final and fourth piece of suggestive evidence in support of our hypotheses.

To address potential concerns about violations of the exclusion restriction we show that it is unlikely that the chain of causality runs in the reverse direction --- that is, that the unconditional cash transfer or the psychotherapy intervention led to the alignment of time preferences between the woman and her partner, and subsequently changed the likelihood of IPV. We also provide suggestive evidence that the treatment did not have a direct effect on time preferences. Finally, we show that the interventions do not appear to have had an effect on time preferences through an indirect channel of stress reduction.

We contribute to the literature on the impact of IPV on economic outcomes of victims. One prior empirical paper suggests that women in abusive relationships have worse economic outcomes due to IPV, such as reduced earnings and employment \citep{Adamsetal:2024}. An explanation, akin to \citet{AnderbergRainer:2013}, is that partners may use threats of violence to sabotage women's labour market choices directly. Another explanation is that threats and acts of violence indirectly influence economic outcomes by changing the victims' preferences. For example, if IPV makes women more impatient, this may prevent women from making necessary long-term investments, e.g. savings decisions, taking-up job opportunities as well as actions needed to exit violent relationships. We put forward novel empirical evidence on this channel.

Secondly, we contribute to a literature on how trauma shapes economic preferences. We study the causal effect of exposure to violence in the household on decision making. The closest to our study is \citet{Callenetal2014} who provide causal evidence of the effect of exposure to violence in public spaces on risk preferences. They find that individuals who have been exposed to violence (e.g., direct fire, improvised explosive device explosions, indirect fire, mine strikes, and suicide attacks), when randomly asked to recall fearful experiences, in some contexts demonstrate risk tolerant behaviour. Other correlational studies show that individuals' risk, time and social preferences are altered after a natural disaster or conflict \citep{Eckeletal:2009, Voorsetal:2012, Cassaretal:2017}. Neuroscience literature also provides correlational evidence indicating that PTSD patients, including women with experiences of IPV, exhibit reduced neuropsychological functioning in decision-making compared to individuals without PTSD \citep{Twamleyetal:2009, Aupperleetal:2016, Steinetal:2002, Scottetal:2015}. Our results 
that if patterns of behaviours can be triggered for women exposed to IPV, then this can be potentially actionable by policy intervention, 
 or even adversely be used strategically by violent partners. 

Finally, we connect to a broader literature on using randomly assigned recall (priming) techniques to study economic behaviour. This technique has been used to study the impact of social identity, violence or poverty on economic outcomes; see \citet{CohnMarechal:2016} for a review. We contribute to this literature by being the first to use this method to estimate the causal impacts of IPV on economic decision making.

\section{Design of randomized recall experiments}
\label{sec:design}
We present the design of the randomized recall experiments in three sub-sections. We first describe the recall methods we use to exogenously manipulate the recollection of IPV exposure. Next we summarize the experimental design, and finally, the sample selection in each of the two recall experiments.

\subsection{Recall of IPV events}
Studies in psychology and economics have developed a series of methodologies for cueing an emotional state or identity. While the method does not manipulate whether an individual experienced an event or not, for a group of individuals randomly allocated to the treatment group they are asked to recollect these experiences. Priming as a technique opens up avenues for causal identification where it is impossible or unethical for the researcher to administer or manipulate exogenously the event itself, e.g., traumatic or violent experiences, poverty, or social identity. Importantly, unlike in other correlational analyses where measurements are taken sometime after exposure to an event, the random recollection generates a sample that is balanced on both exposure and other unobservables. Thus, any differences in preferences or decision making post-recollection between the random recall arm and the placebo arm, are attributed to the emotional state cued by the recall exercise.

We conducted two experiments in Ethiopia using recall of IPV events that individuals had experienced, as a way to causally identify the impact of IPV exposure on time preferences.\footnote{In this first experiment we also tested effects on risk preferences and cognitive functioning.} Just prior to completing tasks to elicit these outcomes, subjects were randomly assigned to one of two experimental arms. Respondents assigned to the treatment arm were asked about their IPV experiences as a method of making any past experience of IPV salient. Our recall method is based on the survey module in the Demographic and Health Survey \citep{CSACE2016} and the WHO Violence Against Women Instrument \citep{Ellsbergetal2001} --- both internationally validated survey instruments that have been included in large-scale national household surveys across multiple developing countries and time periods, and used in economic research on IPV (e.g., \citet{KotsadamVillanger:2022, HaushoferShapiro2016, Alesinaetal:2016,  Greenetal:2020, Anderbergetal2024, Hidroboetal:2016, Heathetal2020}). In other words, a standard IPV module serves as the recall method. The questions present a range of situations asking about experiences (Table \ref{tab:DHS_instrument}) and physical trauma from IPV.\footnote{Given the sensitive nature of IPV, the prime started with a preface: ``\textit{Now I would like to ask you questions about some other important aspects of a woman's life. You may find some of these questions very personal. However, your answers are crucial for helping to understand the condition of women in Ethiopia. Let me assure you that your answers are completely confidential and will not be told to anyone, and no one else in your household will know that you were asked these questions. If I ask you any question you don't want to answer, just let me know and I will go on to the next question.}" This wording was deliberately designed to make the respondent comfortable to report violence, rather than framing it as something shocking or out of the ordinary. This is similar to the DHS, and in line with the WHO protocol on ethical guidelines for conducting research on IPV \citep{WHO2016}. Additionally, the assured confidentiality and interviews being conducted in a safe location contributed to making her feel safe and improve truthful reporting of IPV experiences \citep{Ellsbergetal2001}.}

The other half of subjects were asked a placebo module. In this module the respondent was presented with a set of hypothetical placebo scenarios, after which she answered a few questions. The scenarios and questions were worded in a manner to mimic casual `chit-chat' or small-talk, based on extensive piloting. In the first scenario, the respondent was asked about her favourite item of clothing or shoes or jewellery. She was then asked to elaborate specifics, like the color, why she likes this item, material and how she bought, made or got it. The second scenario covered a celebratory meal with the family in a festive get-together. She was asked to elaborate what she would prepare, ingredients she would purchase and with whom she would prepare the meal. In the third scenario, the respondent was asked about her favourite Ethiopian holiday. She was asked to elaborate what she would do to celebrate that holiday. The modules were piloted such that time taken to administer them was approximately similar.

\subsection{Design}
The randomized recall experiment was conducted via in-person interviews by trained enumerators. The experiment began with the recall exercise that the respondent was randomly assigned to, i.e., the treatment arm (``IPV Recall") or the control arm (``Placebo Recall"). Next, respondents carried out the incentivized activities to measure time preferences. Finally, women in the control arm were also asked the IPV module. This implies that we have data on IPV from all participants, but the primary outcomes did not follow recall of IPV for those women assigned to the control arm.

Randomization was done at the individual-level. In the first experiment, randomization was stratified by before or after pay-day and four employment groups.\footnote{The four employment groups were: never had a formal job or owned a business; currently working in factories; left factory employment and were not working formally or owning a business; left factory employment and were now engaged in non-factory formal employment or running a business} In the second experiment, randomization was stratified by region (Amhara or Oromia), experimental arms of the RCT (cash;  psychotherapy; cash and psychotherapy; or control), and women's cohabitation status as per the previous survey round.

\subsection{Sample}

For the first experiment, we sampled women in urban and semi-urban regions in Ethiopia who were cohabiting with a male partner in the past three months. We sampled women from four employment groups: i) women who were working in factories; ii) women who had worked in factories but subsequently stopped and were now employed elsewhere or self-employed; iii) women who had worked in factories but subsequently stopped, and were now unemployed; iv) women who had never worked in factories, but who now worked and lived in communities which are heavily dependent on factory employment. We used a wait-list design to select an equal proportion from the four employment groups. In particular, women with a randomly-assigned number lower than the remaining targeted number of women were first invited for the experiment; while the remaining women were reserved as potential replacement sample, to be contacted in order of their random number if the initially-selected respondents did not end up participating. The sample for this experiment was 400 women.

In the second experiment, we leveraged the sample of female respondents of an ongoing randomized control trial in rural regions in Ethiopia.\footnote{The ongoing RCT studies the impact of a psychotherapy program with and without a one-time lump sum cash transfer on measures of both economic and psychological well-being outcomes. See \href{https://www.socialscienceregistry.org/trials/10201}{AEARCTR-0010201} for details on the RCT.} The sample consists of women from households that are registered and assisted by the Productive Safety Net Program (PSNP) in Amhara and Oromia regions of Ethiopia. Furthermore, to be selected into the sample for the RCT, individuals needed to present signs and symptoms of depression or dysfunction, be between 18 and 59 years old, and be the main decision maker or the spouse of the main decision maker in the household at baseline. The sample for this experiment was 1,666 women.

\section{Data and descriptive statistics}
\label{sec:data}

\subsection{Time preferences}
We elicited time preferences using multiple price lists over a monetary domain \citep{Cohenetal:2020}.
The respondent was asked to make choices between receiving 50 ETB (Birr) on an earlier date or a differing amount on a later date. There were 16 choices in the first experiment, ranging from 30 Birr to 200 Birr, and 13 choices in the second experiment, ranging from 1 Birr to 250 Birr. The respondent was informed that if the task was randomly selected for payment, she would be paid by cash and be given a written `I owe you' (IOU) on the day of the interview for the amount and payment date. This implies that choosing to take the money today versus on another date would not create any differential risk that she would not be paid. We elicited time preferences for two time horizons: 1) near frame -- 1 day (3 days) or a varying amount in 14 days (10 days), and 2)  far frame -- 14 days or a varying amount in 28 days (21 days). To ensure incentive compatibility, the respondent was informed that if the task was selected during the random lottery for payment, only one randomly-selected frame, and only one randomly-selected choice in that frame would be executed for real payment.

We expect most, but not necessarily all respondents to prefer 50 Birr on the earlier date versus the substantially lower amount on the later date. However, at a certain point, a rational respondent should switch over from the earlier date to the later date, and from then on prefer the amount on the later date. We did not enforce single-switching. However, if a respondent switched multiple times, enumerators were prompted to double-check comprehension of the task.

We construct a continuous variable for each time frame that takes the monetary value of the respondent's first switch-point, i.e., the amount of money on the later date. This provides a common monetary metric across the two survey experiments. By construction, this implies that individuals who switch on a later question demonstrate more impatient behaviour, since they would need a higher amount on the later date in order to switch from preferring 50 Birr on the earlier date. 

We extensively trained the enumerators and piloted the task to ensure comprehension among a low-literacy sample. We test for violations of the task: less than 1\% of respondents switch multiple times. We also test for comprehension of the task with two questions. We first ask: ``\textit{Would you rather receive 60 ETB or 70 ETB?}" A rational individual should choose the higher amount of 70 Birr. We then ask: ``\textit{Would you rather receive 70 ETB in three days or 70 ETB in 10 days?}" A rational individual  should weakly prefer the amount on the earlier date. We find only 3 respondents answered both comprehension question incorrectly.

\subsection{Balance}
\label{sec:bal}
 Table \ref{balance} shows balance across experimental arms in each of the experiments on variables pre-specified in the pre-analysis plan (PAP) and balance on IPV experiences.\footnote{We exclude pre-specified balance variable ``household size'' in the first experiment because this is a variable we construct from three survey questions, namely the number of dependants, the number of children aged 0-5, and the number of children aged 6-18. This implies that these variables are perfectly collinear. We exclude pre-specified balance variable ``no-formal employment of partner'' in the second experiment because less than 7\% of our sample was in this category.} For each variable, we report the means in the treatment arm (T), the means in the control arm (C), and the difference in means ($T-C$)
as well as the corresponding \textit{p}-value for the test that the difference in means is zero, and the normalized differences. Only 1 test out of 22 has a \textit{p-}value of less than 0.10, namely if the woman has a child less than 6 years of age in the second experiment. The normalized differences show that the significant difference in means is small, and close to the rule of thumb of 0.25 as suggested by \citet{ImbensRubin2015}. The \textit{p-}value of the Chi-squared test of joint significance is not significant for the differences in means for each experiment.\footnote{The first experiment had a second treatment arm, where respondents were asked to recall financial concerns in a module similar to \citet{Manietal2013}. Table \ref{balance_ex1} shows balance across all experimental arms. We drop the respondents in this second treatment arm for the purposes of this paper. }

In terms of IPV rates in the past months: In the first  experiment, 55\% of women in the control group have experienced an act of IPV in the past three months. If we decompose this into the standard categorization of violence, 53\% have experienced emotional violence and 19\% physical or sexual violence. In the second experiment, the rates of violence are 33\%, 31\% and 14\% for any IPV, emotional violence and physical or sexual violence in the past 12 months respectively. 

\begin{table}[H]
\begin{center}
\caption{Descriptive statistics and balance}
\resizebox{0.60\textwidth}{!}{\hspace{-3.5cm}\begin{minipage}{\linewidth}
	\input{"ch3_balance_normdiff_ex1_2_12122024_adj.tex"} 
 \end{minipage}}

\begin{tablenotes}		
			\footnotesize
			\item Notes: The table presents descriptive statistics and balance tests for each experiment (Columns 1-4: First experiment; Column 5-8: Second experiment). Columns 1 and 5 show the means of the variable in the  treatment arm (``IPV recall") and Columns 2 and 6 show the means of the variable in the control arm (``Placebo recall").  Columns 3 and 7 show the difference in means between the treatment and control arms. Robust standard errors are indicated in parentheses. Normalized differences are reported in square brackets, calculated as the difference between the sample means of experimental arms divided by the square root of the sum of the sample variances. The last row ``Joint significance (\textit{p}-value)" reports the p-value on the chi-squared test that coefficients and \textit{p}-values from all regressions on balance variables in each experiment are jointly unrelated to the treatment assignment. Stars indicate: *** 1 percent ** 5 percent * 10 percent level of significance.
		\end{tablenotes}\label{balance}

\end{center}
\end{table}

\subsection{Attrition}
\label{sec:attrit}
The attrition rate for the first experiment is zero due to the wait list design. The attrition rate in the second experiment is 9\%. Attrition is not differential between treatment and control arms (point estimate: -1.4 percentage points, \textit{p}-value: 0.327).

\section{Results of randomized recall experiments}
\label{sec:results}
We test for heterogeneous treatment effects by experiences of IPV pooling data from our two experiments, using the following OLS model (See Figure \ref{fig:estimation} for an illustration):

\begin{equation} \label{eqn:OLS_het}
Y_{ise}=\alpha + \beta T_{ise} +  \phi T_{ise}*V_{ise} +  \theta V_{ise} + \eta_{e} + \tau_{s} + \epsilon_{ise}
\end{equation}
The dependent variable $Y_{ise}$ is the outcome on time preferences for individual $i$ in stratum $s$ and experiment $e$. The treatment variable ($T_{ise}$) takes on value one if the individual was randomly assigned to the treatment arm (``IPV Recall"), and takes on value zero if assigned to the control arm (``Placebo Recall"). The variable \(V_{ise}\) represents the respective binary measure of IPV experiences for individual $i$ in strata $s$ and experiment $e$. The coefficient $\beta$ estimates the treatment effect for women who have not experienced violence. Our coefficient of interest is $\beta+\phi$, which estimates the causal effect of recalling IPV experiences for women who have experienced violence.  We include fixed effects for randomization strata $\tau_{s}$ and each experiment $\eta_{e}$. We estimate heteroskedasticity-robust standard errors as randomization was done at the individual level.

Table \ref{IPV_mo_stacked} presents the estimated heterogeneous treatment effects on time preferences elicited in the near frame and far frame for women who have and have not experienced IPV. Time preferences are measured as the present equivalent of the monetary value on the later date at the first switch point, that is, the minimum amount of money the woman is willing to accept at a later date to switch from receiving 50 Birr on the earlier date. Therefore, by construction, a higher present equivalent implies that a respondent is more impatient, as she is willing to forego a larger amount at a later date in order to accept the 50 Birr (smaller amount) on the earlier date. In Columns 1-2, effects are presented for women who have experienced any act of IPV, in Columns 3-4 for emotional violence only, in Columns 5-6 for physical or sexual violence.

In the control group, we find an average present equivalent of 89.19 Birr in the near frame and 88.62 Birr in the far frame. In other words, on average women are willing to give up around 38-40 Birr in return for a payment at an earlier date. 

We find that women who have experienced IPV events in the past months,\footnote{Past months refers to three months in the first experiment and twelve months in the second experiment.} when randomly asked to recall specific acts and injuries from these events, behave more impatiently, than women with IPV experiences who are not asked to recall these IPV events. Decomposing the effects by the standard categorization of violence in the literature --- emotional, and physical and sexual violence, we find that the effects are driven by emotional violence. Point estimates are similar in the near frame (early date: 1 day (3 days) from now) and far frame (early date: 14 days from now), with the effect being significant in the far frame (p-value $<$ 0.10) and marginally insignificant in the near frame (\textit{p-}value $=$ 0.134). Effect sizes suggest that women who have experienced emotional IPV when asked to recall IPV events are willing to forgo 8.5 and 10.9 Birr respectively to receive the amount of money on the earlier day, as compared to women with similar IPV experiences who are not asked to recall IPV events.\footnote{Emotional violence encompasses situations describing threats to physical or sexual harm and controlling behaviour, rather than the physical or sexual act itself.} This amounts to forgoing 9-12\% of the average present equivalent in the control group.

Finally, for women who have not experienced IPV, being randomly primed to recall IPV experiences does not lead to significant impacts on their time preferences, in the near frame or far frame for neither of the different categories of violence. Effects of recall are significantly different between woman who have experienced IPV and women have not experienced IPV in the past months.\footnote{We do not find significant effects of the intervention on risk preferences and cognitive functioning (Table \ref{IPV_risk_cf}). }

We show that results are qualitatively similar to using implied daily discount rates instead of the monetary present equivalents (Table \ref{IPV_mo_delta_stacked}). We calculate implied daily discount rates, using the formula commonly used in the literature: $\delta = \left(\frac{Y}{X}\right)^{\frac{1}{k}}$ where $X$ is the amount on the earlier date (50 Birr), $Y$ is the amount on the later date (varies) and $k$ is the number of days between the earlier and later date.

\begin{table}[H]
\begin{center}
\caption{Effect of IPV in the past months on time preferences}
\begin{footnotesize} 

 \input{"po_time_m_mon_cohab_ex1_2.tex"}
\begin{tablenotes}		
			\footnotesize
			\item Notes: The table presents marginal treatment effects from OLS regressions specified in Equation \ref{eqn:OLS_het}. The dependent variable is the monetary value of the present equivalent of the respondents’ first-switch point in the near frame (Columns 1, 3 \& 5) and far frame (Columns 2, 4 \& 6), as pre-specified. ``IPV Recall" is a binary independent variable that takes on value one if the respondent was randomly assigned to the treatment arm and zero if assigned to the control arm (``Placebo Recall"). ``IPV" are binary independent variables of Any IPV (Columns 1-2), Emotional IPV (Columns 3-4), and Physical/Sexual IPV (Columns 5-6), measured over the past three months in the first experiment and over the past 12 months in the second experiment.  Fixed effects for randomization strata and experiments are included. Robust standard errors are indicated in parentheses. The row ``Recall” presents estimates for $\beta$, the causal effect of recalling IPV experiences for women who have not experienced recent IPV. The row ``Recall + Recall X IPV” presents estimates for $\beta+\phi$, the causal effect of recalling IPV experiences for women who have experienced violence and the corresponding p-value.  The row ``Control Mean" indicates the average present equivalent among all respondents in the control group. Stars indicate: *** 1 percent ** 5 percent * 10 percent level of significance.
		\end{tablenotes}\label{IPV_mo_stacked}
\end{footnotesize}
\end{center}
\end{table}

As a robustness test for the relevance of recall of IPV events, we run the same  regression for woman who have experienced IPV at least once in their lifetime (Table \ref{IPV_mo_stacked_ev}). Compared to experiencing IPV in recent months, lifetime experiences may have occured in the distant past, and may therefore be harder to recall, or the memories may be less vivid. Indeed, we do not find significant effects of being randomly primed to recall acts of IPV for women who have faced IPV in their lifetime.\footnote{The rate of lifetime IPV is 7 percentage points higher in the first experiment and 16 percentage points higher in the second experiment as compared to IPV in the past months. Women in the second experiment have almost double the years of cohabitation as compared to women in the first experiment and are on average older. This could be the reason for the larger discrepancy between the rate of lifetime IPV as compared to IPV in the past months.}

\subsection{Alternative interpretations of effects on time preferences}
As explained in the introduction, the randomized recall experiment provides a way to attempt causal identification of IPV in a world where randomizing IPV is not ethical nor feasible. Unlike correlational designs, the randomized recall preserves balance on both exposure and (un)observed characteristics. While this leads to an internally valid and robust result of the recall exercise on time preferences, concerns remain with the interpretation, specifically whether the observed behavioural responses truly represents the effect of actual trauma. Therefore we now discuss and attempt to rule out potential alternative interpretations of the results.

\subsubsection{Riskiness of future payments}
A potential concern with providing future payments when conducting experiments to elicit time preferences is that respondents may not trust that future payments will be made, and therefore prefer to receive payments now, rather than in the future. A deliberate design feature to counteract this is that all payments were made at a date in the future, ranging from one day post-survey to four weeks in the future. To further ease these concerns, respondents were given reassurance of the payment, including a written and signed IOU stating the amount and date they would receive the payment. Given that individuals were randomized across treatment arms, as long as mistrust creates noise in both arms, it does not confound our findings.

Another potential concern could be that for women who have experienced IPV the priming heightened the uncertainty of payments at dates in the future, especially if women would not like the payment to be made in the presence of a violent partner.\footnote{We know that there could be income hiding or extraction concerns in households, especially when there is IPV.} This argumentation would mean that the woman would prefer payments on an earlier date as she may be unsure about the ``whereabouts" of the partner on a date in the future. That is, women may appear ``impatient'' by choosing the earlier date but we might not be capturing time preferences. It is important to note that the large significant effects we find are in the \textit{far frame} -- where both payment days are relatively far in the future (14 days versus 28 (21 days)). Thereby reducing both potential concerns.

\subsubsection{Liquidity constraints}
Another potential concern with using money-now versus money-later tasks to measure time preferences is that they can be confounded by liquidity constraints. People who are more liquidity constrained now will prefer to receive money now, not because they are impatient, but because they need the earnings from the task. This would imply that we should observe different behaviour before individuals receive a positive income shock versus after a positive income shock. More precisely, if the results from the randomized recall experiment picks up an effect of liquidity constraints rather than time preferences, we should see that people appear more impatient before a positive income shock than after a positive income shock \citep{Carvalhoetal:2016}. 

To provide evidence that it is unlikely that liquidity constraints are driving our results we first use the same argument as above, that we see the strongest treatment effects in the far frame, where liquidity constraints should play less of a role. Second, we provide suggestive empirical tests for this in both experiments. In the first experiment, we exploit quasi-experimental variation in the date of the interview of the time preference elicitation, and the respondent's payday from the factory, when they receive a substantial income flow. This allows us to categorize our respondents into \textit{before} or \textit{after} pay-day respondents, as proxies for respondents with and without liquidity constraints. In the second experiment, we check for impacts in the intervention arm where a random subset of individuals received an unconditional cash transfer. In Table \ref{tab:IPV_ex1_after_payday} and Table \ref{tab:IPV_ex2_cash_arm} we see that even after liquidity constraints are eased, i.e., for those interviewed after pay day in the first experiment and for those receiving an unconditional cash transfer in the second experiment,  our results are robust, thus suggesting that liquidity constraints are unlikely to explain the effects we find. 

\section{Exploiting RCT-Induced IPV Variation}

As explained above, because IPV cannot be ethically or experimentally assigned, we combine evidence from two complementary second-best strategies. The first, presented above, is from randomized recall experiments. While we addressed some concerns with this method, by design we are unable to address the concern that it does not test the effect of real-life, acute experiences of IPV on time preferences. Therefore we use another second-best strategy whereby we exploit exogenous variation in IPV induced by randomized implementation of interventions. We leverage two experiments -- an unconditional cash transfer (Section \ref{scn:quasi_HRS}) and a psychotherapy intervention (Section \ref{scn:quasi_GPM}) -- that were aimed at reducing IPV. While this method does allow us to exploit variation in real-life IPV, through leveraging the randomized interventions as instruments, the validity of the causal identification strategy -- especially violations of the exclusion restriction -- will remains a valid concern, which we will discuss in Section \ref{scn:quasi_ex_restriction}.

\subsection{Estimation strategy}
\label{scn:IV_estimation_strategy}
We estimate the effect of each intervention on IPV using the following specification:
\begin{align*} \label{eqn:first_stage}
V_{iv} &= \pi_0 + \pi_1 T_v + \eta_{iv} & \text{(First stage)}
\end{align*}

The variable \(V_{iv}\) represents the respective measure of IPV experiences for individual $i$ in village $v$. $T_v$ is a binary variable indicating if the respondent's village was assigned to the treatment or control arm. $\eta_{iv}$ is the error term. $T_v$ is uncorrelated with $\eta_{iv}$ by construction. The coefficient of interest for the first stage is $\pi_1$ which is the change in IPV ($V_{iv}$) due to the intervention. Standard errors are clustered at the unit of randomization. We report robust first-stage \textit{F}-statistics based on
\citet{OleaPflueger:2013}, which are robust to clustering and heteroskedasticity.\footnote{In a just-identified case, the \citet{OleaPflueger:2013} efficient \textit{F}-statistic is equivalent to the \citet{KleibergenPaap:2006} \textit{F}-statistic.}

We leverage the reduction in IPV due to the intervention, to estimate the effect on time preferences through an instrumental variables approach. We estimate the second stage with the following specification using two-stage least squares (2SLS):
\begin{align*}
Y_{iv} &= \gamma_0 + \gamma_1 \hat{V}_{iv} + \epsilon_{iv} & \text{(Second stage)}
\end{align*}

Our coefficient of interest is $\gamma_{1}$, which estimates the effect of the predicted reduction in IPV from the first stage on time preferences ($Y_{iv}$). Our hypothesis is that women who experience a reduction in the likelihood of IPV due to the intervention exhibit more patient behaviour. The sign of this effect is opposite to the hypothesized sign for the survey experiments because in the survey experiment women are being asked to recall events of IPV, making IPV experiences more salient while in the experimental analysis, we are assessing interventions that reduce IPV. Other variables are as described above. 
Standard errors are clustered at the unit of randomization.

\subsection{An unconditional cash transfer as instrument}
\label{scn:quasi_HRS}
\textbf{Set-up:} We leverage a randomized control trial of an unconditional cash transfer intervention in Kenya -- \citep{HaushoferShapiro2016} -- that reduced intimate partner violence among women. The sample consists of 1010 cohabiting women for whom data on IPV was collected at endline, 698 of whom are in treatment villages that received the unconditional cash transfer and 312 in control villages. The cash transfer is substantial and amounts to approximately 10\% of average annual household income for the households in the sample. The data on IPV in \citet{HaushoferShapiro2016} is collected using the same set of survey questions as we use in our survey experiments. The time period over which they measure IPV experiences is during the past six months prior to the survey. We construct similar binary variables and indices as in our survey experiment. We also use their data on time preferences which also consists of a money-now versus money-later task in the near frame (tomorrow versus six months) and far frame (six versus twelve months). The time preference task is not identical to ours but is comparable, respondents were asked to make choices between receiving a differing amount of money on the earlier date (0-100 Kenyan Shillings) versus 100 Kenyan Shillings on a later date. Respondents made 11 choices. Since the amount on the later date remains constant, we construct the monetary value of the present equivalent as the amount on the earlier date at the respondents first switch-point. By construction, this implies that individuals who switch on a later question demonstrate more patient behaviour, since they would need a higher amount on an earlier date to switch to the amount on the later date. That is, our hypothesis is $H_a: \gamma_{1}>0$.

\bigskip

\noindent \textbf{First stage:} As also confirmed in \citet{HaushoferShapiro2016}, their cash transfer intervention led to a significant reduction in IPV -- particularly driven by a reduction in physical and sexual violence (Table \ref{tab:first_stage_HRS}). Women in treatment villages were 8 percentage points less likely to experience physical and sexual violence in the past 6 months, a 21\% reduction as compared to the control mean. When assessed using the violence index, defined as the total count of specific violent acts the respondent reported experiencing, we observe a similar reduction. It is important to note the high rates of IPV in this study: in the control group almost 97\% of women report having experienced any form of IPV, with more than 3 specific types of violent acts being reported on average; 90\% report emotional IPV, with 1.7 specific types of acts reported on average; and 38\% report physical or sexual IPV, with 1.4 specific acts on average, in the past 6 months. \footnote{The item driving the high rates of emotional violence is: ``\textit{During the last 6 months did your husband/partner ever threaten to hurt or harm you or someone close to you}." 77\% women in the sample in Kenya respond yes to this situation. This item is one of the more severe forms of emotional violence, thus, we are reluctant to dismiss the measure due to low variation.}

\begin{table}[H]
\begin{center}
\caption{Impact of an unconditional cash transfer on IPV}
 \label{tab:first_stage_HRS}
 \input{"first_stage_vil_hrs_13122024.tex"}  

\begin{tablenotes}		
			\footnotesize
			\item Notes: The table presents estimates of the effect of the unconditional cash transfer on IPV using OLS regressions, based on the data from \citet{HaushoferShapiro2016}. Dependent variables are measured over the past six months and refer to the following categorizations:  Any IPV (Columns 1-2), Emotional IPV (Columns 3-4) and Physical/Sexual IPV (Columns 5-6). The dummy variable is constructed as a binary variable taking on value one if the respondent answered ``yes” to at least one of the specific types of violent acts listed. The index variable comprises an index by summing up the number of specific types of violent acts to which the respondent answered ``yes” to having experienced. Standard errors are clustered at the village level, which is the unit of randomization, and are indicated in parentheses. The row ``Control Mean" indicates the likelihood of IPV in the control group. The row ``Effective F-stat" reports robust first-stage F-statistics based on \citet{OleaPflueger:2013}. Stars indicate: *** 1 percent ** 5 percent * 10 percent level of significance. 
		\end{tablenotes}

\end{center}
\end{table}

\noindent \textbf{Second stage:} Table \ref{tab:2sls_HRS} presents effects of the predicted reduction in IPV, instrumented by the unconditional cash transfer, on time preferences in the near frame and far frame. We find that a reduction of one specific type of act of IPV due to the intervention leads to an increase in patience in the near and far frame, which is significant in the near frame (\textit{p}-value $<$ 0.10) but marginally insignificant in the far frame (\textit{p}-value = 0.125). This implies that a reduction in the number of specific types of IPV acts experienced in the past 6 months makes women willing to forgo 9-12 Kenyan Shillings on the earlier date in order to receive a larger sum of money at the later date, amounting to approximately 20\% of the control group mean. When instrumenting by the dummy for physical/sexual IPV, a reduction from any physical/sexual violence to no physical/sexual violence due to the intervention, leads to an increase in patience in the near and far frame, which is marginally insignificant in both frames (\textit{p}-value is 0.140 and 0.139, respectively). This implies that a full cessation of physical/sexual IPV due to the intervention makes women willing to forgo 67-78 Kenyan Shillings on the earlier date in order to receive a larger sum of money at the later date. The point estimate is substantially larger for the dummy, as here, the variation exploited is a reduction from any phsyical/sexual IPV to no physical/sexual IPV. We also sacrifice power when using the binary variable, because we lose relevant variation, thereby explaining the inability to detect effects as well as lack of convergence in the estimated Anderson-Rubin upper bound of the confidence intervals. 

The \textit{F-}statistic for the first stage is 8.016 for the index of any IPV, 8.303 for the index of physical/sexual IPV, and 5.629 for the dummy of physical/sexual IPV, which are below the threshold where the bias of the two-stage-least-squares estimator is no more than 10\% of the OLS bias (as per the rule of thumb used in \citet{OleaPflueger:2013}). This implies that our coefficients are estimated with a weak instrument. The estimates are, nevertheless, unbiased as the model is just-identified \citep{AngristKolesar:2024}. We show that the estimates from a limited information maximum likelihood estimator, which is median unbiased, are the same as from the 2SLS estimator which further lends confidence to our results (Table \ref{tab:liml_HRS}). In addition, we calculate Anderson-Rubin confidence intervals that provide valid inference under weak instruments, and find that the null hypothesis is similarly rejected for the indices of IPV at a 90\% confidence interval for the near horizon (Columns 1 \& 3).\footnote{This test controls for the probability of incorrectly rejecting the null hypothesis and recovering the true parameter value when the instrument is weak. It is also an efficient test when the model is just-identified \citep{Andrewsetal:2019}.}  Taken together, and in line with recent developments in the literature on weak instruments, our findings are reasonably informative to provide a suggestive test of our hypothesis: women who experience a reduction in IPV exhibit more patient behaviour.

\begin{landscape}
\begin{table}[H]
\caption{Effects of a reduction in IPV, instrumented by an unconditional cash transfer, on time preferences}\label{tab:2sls_HRS}
\begin{center}

 \input{"2sls_treat_vil_hrs_13122024_ar.tex"}  
 \begin{tablenotes}		
			\footnotesize
			\item Notes: The table presents estimates of the effect of a reduction in IPV, instrumented by being randomized to receive an unconditional cash transfer, on time preferences using 2SLS regressions, based on the data from \citet{HaushoferShapiro2016}. The dependent variable is the monetary value of the present equivalent of the respondents’ first-switch point in the near horizon (Columns 1, 3 and 5) and far horizon (Columns 2, 4 and 6). IPV is measured over the past six months and refers to the following categorizations:  Any IPV (Columns 1-2),  and Physical/Sexual IPV (Columns 3-6). The dummy variable is constructed as a binary variable taking on value one if the respondent answered ``yes” to at least one of the specific types of violent acts listed. The index variable comprises a continuous index by summing up the number of specific types of violent acts to which the respondent answered ``yes” to having experienced. P-values from standard errors  clustered at the village level, which is the unit of randomization, are indicated in parentheses.  The row ``Control Mean" indicates the average present equivalent in the control group. The effective F-statistic from robust first-stage F-statistics based on \citet{OleaPflueger:2013} and  90\% confidence intervals using the Anderson-Ruben test are presented as tests for weak instruments.  Stars indicate: *** 1 percent ** 5 percent * 10 percent level of significance.
		\end{tablenotes}
\end{center}
\end{table}
\end{landscape}

\subsection{A psychotherapy intervention as instrument}
\label{scn:quasi_GPM}

\textbf{Set-up}: For the second quasi-experimental analysis, we leverage a randomized control trial of a psychotherapy intervention in Ethiopia. The sample consists of 1630 
 women, of whom 815 are in villages that received psychotherapy and 815 in villages that did not receive the psychotherapy intervention.\footnote{426 women also received a lump-sum cash transfer with the psychotherapy intervention. 450 women received cash only. Since the cash transfer had no effect on IPV --  we pool the psychotherapy arms for this analysis and combine the cash only arm with the arm where women received neither psychotherapy nor cash. Effects are robust to restricting the analysis using the arm with neither psychotherapy nor cash as the comparison group.} The psychotherapy intervention involves behavioural strategies to address both psychological issues (e.g., stress, fear, feelings of helplessness) and practical problems (e.g.,  conflict in the family). We use the data collected at the second endline survey, that is 1.5 years after the intervention was completed. Since the randomized recall  experiment presented above was embedded within this round of data collection, the description and construction of variables on IPV and time preferences is as outlined in Section \ref{sec:data}. In this study a higher present equivalent implies the woman is more impatient, as she is willing to forego a larger amount at a later date in order to accept the 50 Birr (smaller amount) on the earlier date, and vice versa. That is, the hypothesis is $H_a: \gamma_{1}<0$. The construction of outcome variables, the instrument and instrumental variable analysis was pre-specified.
\bigskip

\noindent \textbf{First-stage}: The psychotherapy intervention led to a significant reduction in IPV -- particularly driven by a reduction in emotional violence (Table \ref{tab:first_stage_GPM}). Women in treatment villages were 8 percentage points less likely to experience any violence and 6.6 percentage points less likely to experience emotional violence in the past 12 months, a 21\% reduction as compared to the control mean (\textit{p}-value $<$ 0.05). When measured by the index of violence we find point estimates in the hypothesized direction, however effects are not statistically significant. It is important to reiterate that rates of IPV in the two quasi-experimental studies are quite different. The sample and contexts are also not the same.\footnote{One plausible reason for the difference in rates of IPV is that the Kenya unconditional cash transfer intervention was conducted in 2011, almost 12 years prior to the Ethiopia psychotherapy intervention. There have been rapid declines worldwide in the rates of IPV since then. In addition, national rates of IPV in Ethiopia are lower than Kenya as per the 2015-2016 DHS \citep{UN_Ethiopia:2024, UN_Kenya:2024}.}

\begin{table}[H]
\begin{center}
\caption{Impact of psychotherapy intervention on IPV}
\label{tab:first_stage_GPM}
 \input{"first_stage_allarms.tex"}  
\begin{tablenotes}		
			\footnotesize
			\item Notes: The table presents estimates of the effect of a psychotherapy intervention on IPV using OLS regressions as pre-specified.  Dependent variables are measured over the past twelve months and refer to the following categorizations:  Any IPV (Column 1-2), Emotional IPV (3-4) and Physical/Sexual IPV (Column 5-6). The dummy variable is constructed as a binary variable taking on value one if the respondent answered ``yes” to at least one of the specific types of violent acts listed. The index variable comprises a continuous index by summing up the number of specific types of violent acts to which the respondent answered ``yes” to having experienced.  Standard errors clustered at the sub-district level, which is the unit of randomization, are indicated in parentheses. The row ``Control Mean" indicates the likelihood of IPV in the control group. The effective F-statistic from robust first-stage F-statistics based on \citet{OleaPflueger:2013} is presented. Stars indicate: *** 1 percent ** 5 percent * 10 percent level of significance.
		\end{tablenotes}
\end{center}
\end{table}

\noindent \textbf{Second stage}: Table \ref{tab:2sls_GPM} presents effects on time preferences in the near frame and far frame. We find that a reduction in IPV, from any IPV to no IPV, or from any emotional IPV to no emotional IPV, leads to negative point estimates on time preferences. The effect is marginally insignificant in the far frame (\textit{p}-value is 0.119 and 0.149, respectively). The negative point estimate suggests that women demonstrate more patient behaviour, in line with our hypothesis. A full removal of any IPV and Emotional IPV due to the intervention makes women willing to wait for 58-88 Birr more on the later date rather than accepting the 50 Birr (smaller amount) on the earlier date. In the estimation, statistical power is low due to the use of a dummy where we lose relevant variation, and end up with weak instruments. Since the analysis was pre-specified, we take the correct direction of the point estimates -- despite them being marginally insignificant -- as a final and fourth piece of suggestive evidence supporting our hypothesis that women who experience a reduction from any IPV or emotional IPV to no IPV or no emotional IPV, exhibit more patient behaviour. We are unable to present Anderson-Rubin confidence intervals as they do not converge since effects are not significant at conventional levels of statistical significance.

\begin{table}[H]
\caption{Effects of a reduction in IPV, instrumented by a psychotherapy intervention, on time preferences}\label{tab:2sls_GPM}
\begin{center}

 \input{"2sls_treat_vil_allarms.tex"}
 \begin{tablenotes}		
			\footnotesize
			\item Notes: The table presents estimates of the effect of a reduction in IPV, instrumented by being randomized to receive psychotherapy, on time preferences using 2SLS regressions as pre-specified. Data is from the psychotherapy RCT. The dependent variable is the monetary value of the present equivalent of the respondents’ first-switch point in the near horizon (Columns 1 and 3) and far horizon (Columns 2 and 4). IPV is measured over the past twelve months and refers to the following categorizations:  Any IPV (Columns 1-2), and Emotional IPV (Columns 3-4). The dummy variable is constructed as a binary variable taking on value one if the respondent answered ``yes” to at least one of the specific types of violent acts listed. P-values based on  standard errors clustered at the sub-district level, which is the unit of randomization, are indicated in parentheses. The row ``Control Mean" indicates the average present equivalent in the control group. The effective F-statistic from robust first-stage F-statistics based on \citet{OleaPflueger:2013}.  Stars indicate: *** 1 percent ** 5 percent * 10 percent level of significance.
		\end{tablenotes}
\end{center}
\end{table}

\subsection{Discussion on the exclusion restriction}
\label{scn:quasi_ex_restriction}
As discussed above, both policy interventions are exogenous by construction, given the random assignment, but concerns about the exclusion restriction remain. While we cannot definitively rule out any of these potential violations of the exclusion restriction, we provide a discussion and suggestive evidence to support its validity.

\bigskip

\noindent \textbf{Reverse causality:} One potential channel through which the exclusion restriction may be violated is through reverse causality. The treatment may have first changed time preferences, thereby potentially making partners' preferences more aligned. \citet{Anderbergetal2024} show that differences in preferences between spouses may be a source of IPV. Thus, alignment of preferences due to the intervention could in turn reduce IPV. For this to be the case we would at the minimum require there to be a positive correlation between the difference between partners' time preferences and IPV. We descriptively investigate this using the \citet{HaushoferShapiro2016} data where time preferences were elicited from both the male and the female partner in the household. We find that there is no correlation between the difference in time preferences of partners, and the likelihood of IPV in the control group (\textit{p-}value $=$ 0.831).\footnote{We cannot perform this analysis for the psychotherapy intervention as time preferences were only collected from female respondents.} Thus, we believe it is unlikely that the chain of causality runs in the reverse direction, that is, the treatment led to the alignment of time preferences, and subsequently changed the likelihood of IPV.

\bigskip

\noindent \textbf{Direct effect}: Another channel, that is closely related to the above, could be that the treatment has a direct effect on time preferences and not through the channel of a reduction in IPV. To support this line of reasoning, under certain assumptions, we should also observe a change in time preferences caused by the treatment for those women that did not experience IPV. To check this, we leverage the data from the psychotherapy intervention, where we have IPV data at baseline prior to the intervention. We find that there is no significant effect of the intervention on the time preferences of women who do not experience any IPV at baseline. Point estimates are small and close to zero (Table \ref{tab:exclusion_GPM}).\footnote{We cannot perform this analysis for the unconditional cash transfer intervention as \citet{HaushoferShapiro2016} do not collect baseline data on IPV from the control group.} Thus, the evidence we have suggests that the intervention does not have a direct effect on time preferences.

\noindent \textbf{Indirect effect}: Another potential violation of the exclusion restriction could stem from an effect of the intervention on time preferences, that did not happen through changes in the likelihood or intensity of IPV, but through another confounding variable. If the unconditional cash transfer or psychotherapy intervention reduced, for example, stress for women these could both explain effects on time preferences and on IPV. In both experiments, data was collected on women's self-reported stress. First, we check whether there is an effect of the interventions on stress. For the unconditional cash transfer intervention in Table \ref{tab:exclusion_stress_HRS}, we find  a significant 0.155 SD reduction in stress (p-value $<$ 0.10). For the psychotherapy intervention in Table \ref{tab:exclusion_stress_GPM}, we find no significant effect on stress. It is important to note that the reduction in stress does not per se violate the exclusion restriction. We know from extensive literature that IPV is a traumatic event and is associated with heightened levels of stress. If the intervention reduced IPV-related stress, and this is the effect we are picking up it is not problematic. A concern with the exclusion restriction arises if women with IPV are also more likely to experience non-IPV related stress, i.e., women who experience IPV are also women who experience the reduction in this type of stress, and the reduction, in turn, drives the changes in time preferences.

For the unconditional cash transfer intervention we do find a significant reduction in stress and we use a similar IV approach as outlined in Section \ref{scn:IV_estimation_strategy}. We test if the change in time preferences due to a reduction in stress resulting from the treatment is comparable to the effects we found for a reduction in IPV. Even though we find that the reduction in stress due to the unconditional cash transfer intervention does not have a significant effect on time preferences, the point estimates are noisily estimated but are of the same magnitude as the effects of IPV (Table \ref{tab:exclusion_HRS}). This implies that we cannot rule out that stress could be a channel through which the exclusion restriction is violated in this experiment. For the psychotherapy intervention there is no effect on stress, while there is an effect on IPV, ruling out stress as a channel through which the exclusion restriction is potentially violated.

\section{Conclusion}
Despite one is three women experiencing intimate partner violence, little is known about the consequences of experiencing IPV on women's decision-making. We know that IPV is a traumatic event, and we know from other contexts that trauma can affect decision making. In this study, we ask to what extent violence events in one's own household, in particular intimate partner violence, alters women's preferences. The results from our randomized recall experiments and instrumental variables approaches provide consistent, novel causal evidence that exposure to IPV induces individuals to discount the future more heavily. This evidence suggests a psychological channel through which violence can perpetuate economic disadvantage and constrain women’s ability to take actions. If IPV affects women's time preferences this may influence their long-term investments, e.g., savings decisions, taking-up of job opportunities as well as actions needed to exit violent relationships. Another implication from our results is that trauma can lead to short-term effects on preferences, as opposed to the suggestion in correlational studies that preferences are forever altered post-event. If patterns of behaviours can be triggered for women exposed to IPV, then this can be potentially actionable by policy interventions that are aimed at reducing IPV, or even adversely be used instrumentally by violent partners.

\bibliography{bibliography}



\appendix
\setcounter{table}{0}
\setcounter{figure}{0}
\renewcommand{\thetable}{A\arabic{table}}
\section{Appendix}
\label{scn:appendix}

\begin{figure}[H]
\caption{Illustration of experimental design and estimation strategy}
\includegraphics[scale=1.2]{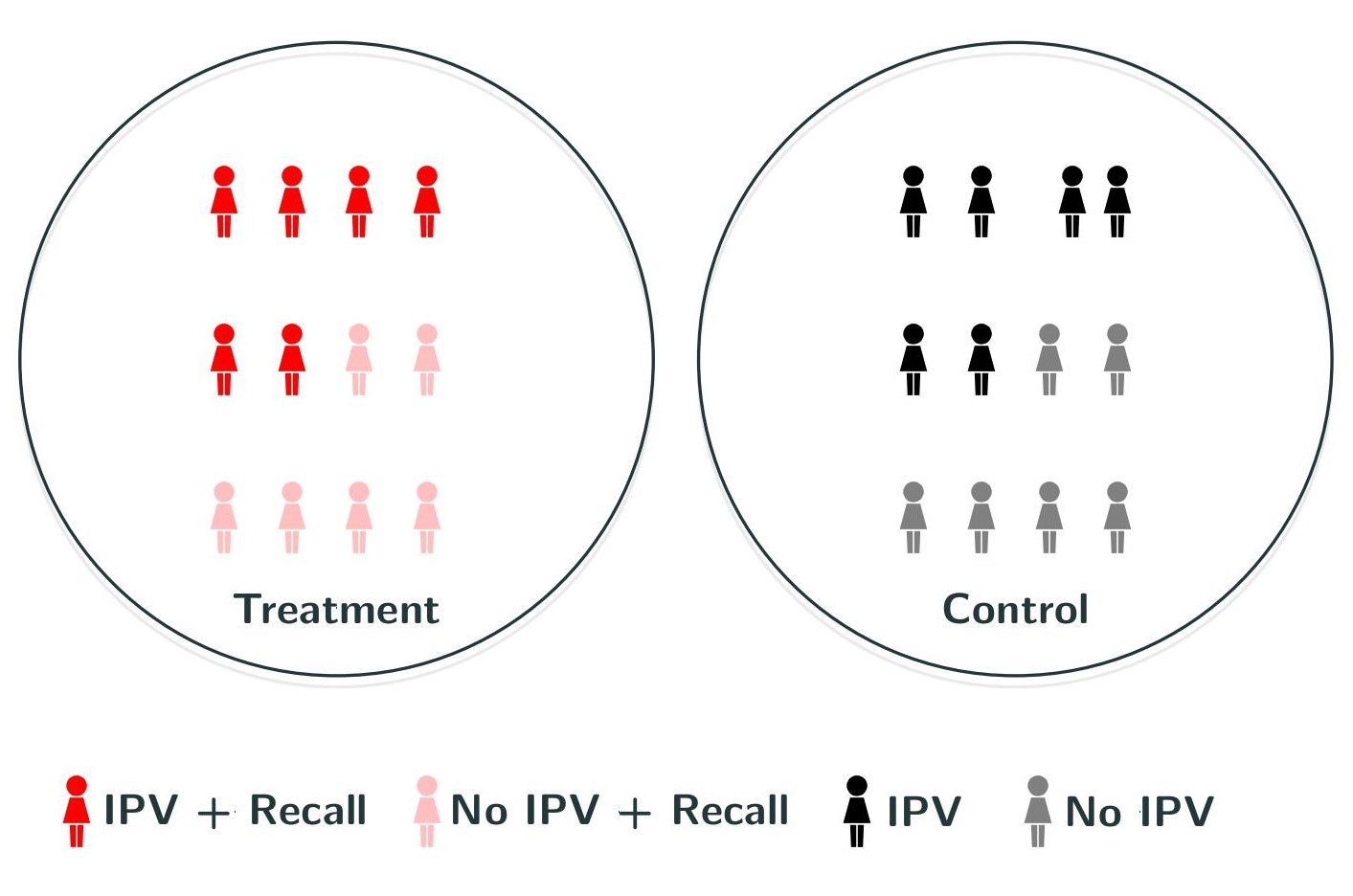}\label{fig:estimation}
\begin{justify}
\begin{footnotesize}
Notes: The figure presents an illustration of the experimental design and estimation strategy for the survey experiments. The left circle with red figures represent women randomized to the treatment arm ``IPV Recall" and the right circle with black figures represent women randomized to the control arm ``Placebo Recall". Both arms contain women who have experienced IPV in past months (darker color markers) and those who have not experienced IPV in the past months (lighter coloured markers). Estimates for $\beta$, the causal effect of recalling IPV experiences for women who have not experienced violence, is estimated by comparing the lighter coloured red markers to black markers. Estimates for $\beta+\phi$, the causal effect of recalling IPV experiences for women who have experienced violence, is estimated by comparing the darker coloured red markers to black markers.
\end{footnotesize}
\end{justify}
\end{figure}

\newpage
\begin{table}[H]
\caption{DHS and WHO survey instrument on intimate partner violence} \label{tab:DHS_instrument}
\begin{longtable}{|p{0.7cm}|p{12cm}|p{2cm}|}
\hline
\textbf{No.} & \textbf{Item} & \textbf{Category} \\ \hline
\endhead
1 & Does not permit you to meet your female friends? & EV \\ \hline
2 & Tries to limit your contact with your family? & EV \\ \hline
3 & Insists on knowing where you are at all times? & EV \\ \hline
4 & Gets jealous or angry if you talk to other men? & EV \\ \hline
5 & Frequently accuses you of being unfaithful? & EV \\ \hline
6 & Insult you or make you feel bad about yourself? & EV \\ \hline
7 & Say or do something to humiliate you in front of other people? & EV \\ \hline
8 & Threaten to hurt or harm you or someone you care about? & EV \\ \hline
9 & Slap you? & PV \\ \hline
10 & Push you, shake you, or throw something at you? & PV \\ \hline
11 & Twist your arm or pulled your hair? & PV \\ \hline
12 & Punch you with his fist or with something else that could hurt you? & PV \\ \hline
13 & Kick you, dragged you or beat you up? & PV \\ \hline
14 & Try to choke or burn you on purpose? & PV \\ \hline
15 & Threaten or attack you with a gun, knife or other weapon & PV \\ \hline
16 & Physically force you to have sexual intercourse with him when you did not want to? & SV \\ \hline
17 & Physically force you to perform any other sexual acts you did not want to? & SV \\ \hline
18 & Force you with threats or in any other way to perform sexual acts you did not want to? & SV \\ \hline
\end{longtable}
\begin{tablenotes}
\footnotesize \item Notes: The Table presents items and categorization of the DHS and WHO instrument on intimate partner violence. Women who are cohabiting with an intimate partner are asked for each act whether she has experienced this act in the past months, and whether she experienced this in her lifetime. The instrument is used in global IPV research to benchmark effects across time periods and contexts. EV refers to emotional violence, PV refers to physical violence and SV refers to sexual violence.
\end{tablenotes}
\end{table}

\begin{table}[H]
\begin{center}
\caption{Balance checks for first survey experiment}
\begin{footnotesize}
\input{"balance_ex1_normdiff.tex"} 
\begin{tablenotes}		
			\footnotesize
			\item  Notes: The table presents balance tests for the first experiment for the two survey experiment treatment arms -- ``IPV recall" and ``Financial Concerns Recall" -- and the control arm -- ``Placebo recall". The difference in means are presented for each comparison. In parentheses are robust standard errors based on individual-level randomization. Normalized differences are reported in square brackets, calculated as the difference between the sample means of experimental arms divided by the square root of the sum of the sample variances. The last row ``Joint significance (\textit{p}-value)" reports the \textit{p}-value on the chi-squared test that coefficients and \textit{p}-values from all regressions on balance variables in each experiment are jointly unrelated to the treatment assignment. Stars indicate: *** 1 percent ** 5 percent * 10 percent level of significance.
		\end{tablenotes}\label{balance_ex1}
\end{footnotesize}
\end{center}
\end{table}

\begin{landscape}
\begin{table}[H]
\begin{center}
\caption{Effect of IPV in the past months on time preferences (excluding multiple switchers).}
\begin{footnotesize}
 \input{"po_time_m_mon_cohab_mswitch_ex1_2.tex"} 
\begin{tablenotes}		
			\footnotesize
			\item  Notes: The table presents marginal treatment effects from OLS regressions specified in Equation \ref{eqn:OLS_het}, excluding the respondents who violated the single switching. The dependent variable is the monetary value of the present equivalent of the respondents' first-switch point in the near frame (Columns 1, 3 \& 5) and far frame (Columns 2, 4 \& 6), as pre-specified. ``IPV Recall" is a binary independent variable that takes on value one if the respondent was randomly assigned to the treatment arm and zero if assigned to the control arm (``Placebo Recall"). ``IPV" are binary independent variables of Any IPV (Columns 1-2), Emotional IPV (Columns 3-4), and Physical/Sexual IPV (Columns 5-6), measured over the past three months in the first experiment and over the past 12 months in the second experiment.  Fixed effects for randomization strata and experiments are included. Robust standard errors are indicated in parentheses. The row ``Recall” presents estimates for $\beta$, the causal effect of recalling IPV experiences for women who have not experienced violence. The row ``Recall + Recall X IPV” presents estimates for $\beta+\phi$, the causal effect of recalling IPV experiences for women who have experienced violence and the corresponding p-value. The row ``Control Mean" indicates the average present equivalent in the control group. Stars indicate: *** 1 percent ** 5 percent * 10 percent level of significance.
		\end{tablenotes}\label{IPV_psv_stacked_switch}
\end{footnotesize}
\end{center}
\end{table}
\end{landscape}

\begin{table}[H]
\begin{center}
\caption{Effect of IPV in the past months on time preferences (in discount factor)}
 \label{IPV_mo_delta_stacked}
  \input{"po_time_m_mon_discount_cohab_ex1_2.tex"}

\begin{tablenotes}		
			\footnotesize
			\item Notes: The table presents marginal treatment effects from OLS regressions specified in Equation \ref{eqn:OLS_het}. The dependent variable is the implied daily discount rates of the respondents’ first-switch point, using the formula $\delta = \left(\frac{Y}{X}\right)^{\frac{1}{k}}$ where $X$ is the amount on the earlier date (50 Birr), $Y$ is the amount on the later date (varies) and $k$ is the number of days between the earlier and later date, in the near horizon (Column 1, 3 \& 5) and far horizon (Column 2, 4 \& 6), as pre-specified. ``IPV Recall" is a binary independent variable that takes on value one if the respondent was randomly assigned to the treatment arm and zero if assigned to the control arm (``Placebo Recall"). ``IPV" are binary independent variables of Any IPV (Columns 1-2), Emotional IPV (Columns 3-4), and Physical/Sexual IPV (Columns 5-6), measured over the past three months in the first experiment and over the past 12 months in the second experiment.  Fixed effects for randomization strata and experiments are included. Robust standard errors are indicated in parentheses. The row ``Recall” presents estimates for $\beta$, the causal effect of recalling IPV experiences for women who have not experienced violence. The row ``Recall + Recall X IPV” presents estimates for $\beta+\phi$, the causal effect of recalling IPV experiences for women who have experienced violence and the corresponding p-value. The corresponding p-value is also presented.  The row ``Control Mean" indicates the average present equivalent in the control group. Stars indicate: *** 1 percent ** 5 percent * 10 percent level of significance.
		\end{tablenotes}

\end{center}
\end{table}

\begin{table}[H]
\begin{center}

\caption{Effect of IPV in one's lifetime on time preferences} \label{IPV_mo_stacked_ev}
\input{"po_time_m_mon_ever_cohab_ex1_2.tex"}
\begin{tablenotes}		
			\footnotesize
			\item Notes: The table presents marginal treatment effects from OLS regressions specified in Equation \ref{eqn:OLS_het}. The dependent variable is the monetary value of the present equivalent of the respondents’ first-switch point in the near horizon (Column 1, 3 \& 5) and far horizon (Column 2, 4 \& 6), as pre-specified. ``IPV Recall" is a binary independent variable that takes on value one if the respondent was randomly assigned to the treatment arm and zero if assigned to the control arm (``Placebo Recall"). ``IPV" are binary independent variables of Any IPV (Columns 1-2), Emotional IPV (Columns 3-4), and Physical/Sexual IPV (Columns 5-6), measured over the women's lifetime. Fixed effects for randomization strata and experiments are included. Robust standard errors are indicated in parentheses. The row ``Recall” presents estimates for $\beta$, the causal effect of recalling IPV experiences for women who have not experienced violence. The row ``Recall + Recall X IPV” presents estimates for $\beta+\phi$, the causal effect of recalling IPV experiences for women who have experienced violence in their lifetime and the corresponding p-value. The row ``Control Mean" indicates the average present equivalent in the control group. Stars indicate: *** 1 percent ** 5 percent * 10 percent level of significance.
		\end{tablenotes}

\end{center}
\end{table}

\begin{table}[H]
\begin{center}
\caption{Effect of IPV on risk preferences and cognitive functioning} \label{IPV_risk_cf}

\resizebox{0.65\textwidth}{!}{\hspace{-1.5cm}\begin{minipage}{\linewidth}
\input{"po_risk_cf_miss_cohab_ex1_2.tex"}  
 \end{minipage}}

\begin{tablenotes}		
			\footnotesize
			\item Notes: The table presents marginal treatment effects from OLS regressions specified in Equation \ref{eqn:OLS_het}. In Columns 1, 3 \& 5, the dependent variable for risk preferences is the monetary value of the respondent’s first switch-point (i.e., their lowest present equivalent compared to the lottery). In Columns 2, 4 \& 6, the dependent variable for cognitive functioning is the length of the sequence of digits that the respondent recalled correctly on the digit span task. These are as pre-specified. ``IPV Recall" is a binary independent variable that takes on value one if the respondent was randomly assigned to the treatment arm and zero if assigned to the control arm (``Placebo Recall"). ``IPV" are binary independent variables of Any IPV (Columns 1-2), Emotional IPV (Columns 3-4), and Physical/Sexual IPV (Columns 5-6), measured over the past three months in the first experiment and over the past 12 months in the second experiment.  Fixed effects for randomization strata and experiments are included. Robust standard errors are indicated in parentheses. The row ``Recall” presents estimates for $\beta$, the causal effect of recalling IPV experiences for women who have not experienced violence. The row ``Recall + Recall X IPV” presents estimates for $\beta+\phi$, the causal effect of recalling IPV experiences for women who have experienced violence and the corresponding p-value. The row ``Control Mean" indicates the average present equivalent in the control group. Stars indicate: *** 1 percent ** 5 percent * 10 percent level of significance.
		\end{tablenotes}

\end{center}
\end{table}

\begin{table}[H]
\begin{center}
\caption{Effect of IPV in the past months on time preferences after payday in the first experiment}
\begin{footnotesize}
 \input{"po_time_m_mon_cohab_ex1_before0.tex"} 
 \begin{tablenotes}		
			\footnotesize
			\item Notes: The table presents marginal treatment effects from OLS regressions specified in Equation \ref{eqn:OLS_het} restricting the sample to the respondents who were interviewed after the payday of the factory. Data is from experiment 1 only. The dependent variable is the monetary value of the present equivalent of the respondents’ first-switch point in the near horizon (Columns 1 \& 3) and far horizon (Columns 2 \& 4), as pre-specified. ``IPV Recall" is a binary independent variable that takes on value one if the respondent was randomly assigned to the treatment arm and zero if assigned to the control arm (``Placebo Recall"). ``IPV" are binary independent variables of Any IPV (Columns 1-2) and Emotional IPV (Columns 3-4), measured over the past three months.  Fixed effects for randomization strata and experiments are included. Robust standard errors are indicated in parentheses. The row ``Recall” presents estimates for $\beta$, the causal effect of recalling IPV experiences for women who have not experienced violence. The row ``Recall + Recall X IPV” presents estimates for $\beta+\phi$, the causal effect of recalling IPV experiences for women who have experienced violence and the corresponding p-value. The row ``Control Mean" indicates the average present equivalent in the control group. Stars indicate: *** 1 percent ** 5 percent * 10 percent level of significance.
		\end{tablenotes}\label{tab:IPV_ex1_after_payday}
\end{footnotesize}
\end{center}
\end{table}

\begin{table}[H]
\begin{center}
\caption{Effect of IPV in the past months on time preferences for the subgroup that received a cash transfer in the second experiment}
\begin{footnotesize}
 \input{"po_time_m_mon_cohab_ex2_cash_only.tex"} 
\begin{tablenotes}		
			\footnotesize
			\item Notes: The table presents marginal treatment effects from OLS regressions specified in Equation \ref{eqn:OLS_het} restricting the sample to the respondents who were in the experimental arms of the RCT that received an unconditional cash transfer. Data is from experiment 2 only. The dependent variable is the monetary value of the present equivalent of the respondents’ first-switch point in the near horizon (Column 1 \& 3) and far horizon (Column 2 \& 4), as pre-specified. ``IPV Recall" is a binary independent variable that takes on value one if the respondent was randomly assigned to the treatment arm and zero if assigned to the control arm (``Placebo Recall"). ``IPV" are binary independent variables of Any IPV (Columns 1-2) and Emotional IPV (Columns 3-4), measured over the past 12 months in the second experiment.  Fixed effects for randomization strata and experiments are included. Robust standard errors are indicated in parentheses. The row ``Recall” presents estimates for $\beta$, the causal effect of recalling IPV experiences for women who have experienced violence. The row ``Recall + Recall X IPV” presents estimates for $\beta+\phi$, the causal effect of recalling IPV experiences for women who have experienced violence and the corresponding p-value. The row ``Control Mean" indicates the average present equivalent in the control group. Stars indicate: *** 1 percent ** 5 percent * 10 percent level of significance.
		\end{tablenotes}\label{tab:IPV_ex2_cash_arm}
\end{footnotesize}
\end{center}
\end{table}

\begin{table}[H]
\caption{Effects of a reduction in IPV, instrumented by an unconditional cash transfer, on time preferences (LIML)}\label{tab:liml_HRS}
\begin{center}

 \input{"liml_treat_vil_hrs_13122024.tex"}  
  \begin{tablenotes}		
			\footnotesize
			\item Notes: The table presents estimates of the effect of a reduction in IPV, instrumented by being randomized to receive an unconditional cash transfer, on time preferences using LIML regressions, based on the data from \citet{HaushoferShapiro2016}. The dependent variable is the monetary value of the present equivalent of the respondents’ first-switch point in the near horizon (Columns 1, 3 and 5) and far horizon (Columns 2, 4 and 6). IPV is measured over the past six months and refers to the following categorizations:  Any IPV (Columns 1-2),  and Physical/Sexual IPV (Columns 3-6). The dummy variable is constructed as a binary variable taking on value one if the respondent answered ``yes” to at least one of the specific types of violent acts listed. The index variable comprises a continuous index by summing up the number of specific types of violent acts to which the respondent answered ``yes” to having experienced. P-values from standard errors are clustered at the village level, which is the unit of randomization, and are indicated in parentheses.  The row ``Control Mean" indicates the average present equivalent in the control group. The effective F-statistic from robust first-stage F-statistics based on \citet{OleaPflueger:2013} and  90\% confidence intervals using the Anderson-Ruben test are presented as tests for weak instruments.  Stars indicate: *** 1 percent ** 5 percent * 10 percent level of significance.
		\end{tablenotes}

\end{center}
\end{table}


\begin{table}[H]
\caption{Effects of a psychotherapy intervention on time preferences for women who did not experience IPV at baseline}\label{tab:exclusion_GPM}
\begin{center}

 \input{"exclusion_directtime_noIPVBL_GPM_allarms.tex"}  
 \begin{tablenotes}		
			\footnotesize
			\item Notes: The table presents estimates of the effect of a psychotherapy intervention on time preferences using OLS regressions for the sample of women who do not experience any IPV in the 12 months prior to baseline. The dependent variable is the monetary value of the present equivalent of the respondents’ first-switch point in the near frame (Columns 1) and far frame (Columns 2). P-values based on  standard errors clustered at the sub-district level, which is the unit of randomization, are indicated in parentheses. The row ``Control Mean" indicates the average present equivalent in the control group. Stars indicate: *** 1 percent ** 5 percent * 10 percent level of significance.
		\end{tablenotes}
\end{center}
\end{table}

\begin{table}[H]
\caption{Effects of a psychotherapy intervention on stress}\label{tab:exclusion_stress_GPM}
\begin{center}

 \input{"exclusion_stress_GPM_allarms.tex"}  
 \begin{tablenotes}		
			\footnotesize
			\item Notes: The table presents estimates of the effect of a psychotherapy intervention on a respondent’s self-reported stress using OLS regressions. The dependent variable is the continuous variable measuring stress using the Perceived Stress Scale, the variable is standardized using the control group mean. P-values based on  standard errors clustered at the sub-district level, which is the unit of randomization, are indicated in parentheses. The row ``Control Mean" indicates the average stress index in the control group, which is close to zero due to standardization. Stars indicate: *** 1 percent ** 5 percent * 10 percent level of significance.
		\end{tablenotes}
\end{center}
\end{table}

\begin{table}[H]
\caption{Effects of an unconditional cash transfer intervention on stress}\label{tab:exclusion_stress_HRS}
\begin{center}

 \input{"exclusion_stress_HRS.tex"}  
 \begin{tablenotes}		
			\footnotesize
			\item Notes: The table presents estimates of the effect of an unconditional cash transfer intervention on a respondent’s self-reported stress using OLS regressions. This is based on the data from \citet{HaushoferShapiro2016}. The dependent variable is the continuous variable measuring stress using the Perceived Stress Scale, the variable is standardized using the control group mean. P-values based on  standard errors clustered at the village level, which is the unit of randomization, are indicated in parentheses. The row ``Control Mean" indicates the average stress index in the control group, which is close to zero due to standardization. Stars indicate: *** 1 percent ** 5 percent * 10 percent level of significance.
		\end{tablenotes}
\end{center}
\end{table}

\begin{table}[H]
\caption{Effects of a reduction in stress, instrumented by an unconditional cash transfer, on time preferences}\label{tab:exclusion_HRS}
\begin{center}
 \input{"2sls_stress_HRS.tex"}  
 \begin{tablenotes}		
			\footnotesize
			\item The table presents estimates of the effect of a reduction in stress, instrumented by being randomized to receive an unconditional cash transfer, on time preferences using 2SLS regressions, based on the data from \citet{HaushoferShapiro2016}. Stress is measured using the Perceived Stress Scale, the variable is standardized using the control group mean. P-values from standard errors are clustered at the village level, which is the unit of randomization, and are indicated in parentheses.  The row ``Control Mean" indicates the average present equivalent in the control group. The effective F-statistic from robust first-stage F-statistics based on \citet{OleaPflueger:2013} are presented.  Stars indicate: *** 1 percent ** 5 percent * 10 percent level of significance.
		\end{tablenotes}
\end{center}
\end{table}

\end{spacing}
\end{document}

%% file: ch3_balance_normdiff_ex1_2_12122024_adj.tex
\begin{tabular}{lcccccccc}
\hline \noalign{\smallskip} &  \multicolumn{4}{c}{Experiment 1}  &  \multicolumn{4}{c}{Experiment 2}  \\\cmidrule(lr){2-5}\cmidrule(lr){6-9}
 & T & C & T-C & N & T & C & T-C & N\\
& (1) & (2) & (3) & (4) & (5) & (6) & (7) & (8)\\
\noalign{\smallskip}\hline \noalign{\smallskip}Age & 31.313 & 30.450 & 0.864 & 263 & 35.019 & 34.897 & 0.122 & 1,666\\
 &   (7.902)   &   (6.366)   &   (0.869)   &      &   (9.654)   &   (9.260)   &   (0.435)   &     \\
 &      &      &   [0.120]   &      &      &      &   [0.013]   &     \\
\noalign{\smallskip}Years of education & 10.007 & 10.109 & -0.101 & 263 & 1.489 & 1.496 & -0.007 & 1,644\\
 &   (3.597)   &   (3.679)   &   (0.433)   &      &   (2.806)   &   (2.826)   &   (0.136)   &     \\
 &      &      &   [-0.028]   &      &      &      &   [-0.003]   &     \\
\noalign{\smallskip}Children less than 6 & 0.724 & 0.791 & -0.067 & 263 & 0.825 & 0.857 & -0.032 & 1,523\\
 &   (0.449)   &   (0.408)   &   (0.053)   &      &   (0.380)   &   (0.350)   &   (0.018)*   &     \\
 &      &      &   [-0.156]   &      &      &      &   [-0.087]   &     \\
\noalign{\smallskip}Children more than 6 & 0.530 & 0.558 & -0.028 & 263 & 0.690 & 0.675 & 0.015 & 1,523\\
 &   (0.501)   &   (0.499)   &   (0.062)   &      &   (0.463)   &   (0.469)   &   (0.023)   &     \\
 &      &      &   [-0.057]   &      &      &      &   [0.032]   &     \\
\noalign{\smallskip}Years cohabitated & 9.188 & 8.881 & 0.307 & 263 & 17.368 & 17.510 & -0.141 & 1,155\\
 &   (8.444)   &   (6.930)   &   (0.946)   &      &   (9.081)   &   (8.917)   &   (0.520)   &     \\
 &      &      &   [0.040]   &      &      &      &   [-0.016]   &     \\
\noalign{\smallskip}Age (partner) & 37.545 & 37.109 & 0.436 & 263 & 41.153 & 41.567 & -0.415 & 1,203\\
 &   (9.707)   &   (8.027)   &   (1.061)   &      &   (11.147)   &   (10.871)   &   (0.637)   &     \\
 &      &      &   [0.049]   &      &      &      &   [-0.038]   &     \\
\noalign{\smallskip}Years of education (partner) & 10.858 & 11.202 & -0.343 & 263 & 2.537 & 2.414 & 0.122 & 1,203\\
 &   (4.459)   &   (4.296)   &   (0.543)   &      &   (3.303)   &   (3.255)   &   (0.189)   &     \\
 &      &      &   [-0.078]   &      &      &      &   [0.037]   &     \\
\noalign{\smallskip}Married & 0.948 & 0.915 & 0.033 & 263 & 0.744 & 0.739 & 0.005 & 1,644\\
 &   (0.223)   &   (0.280)   &   (0.032)   &      &   (0.437)   &   (0.439)   &   (0.006)   &     \\
 &      &      &   [0.130]   &      &      &      &   [0.011]   &     \\
\noalign{\smallskip} Any IPV & 0.463 & 0.558 & -0.095 & 263 & 0.340 & 0.334 & 0.006 & 1,158\\
 &   (0.500)   &   (0.499)   &   (0.061)   &      &   (0.474)   &   (0.472)   &   (0.028)   &     \\
 &      &      &   [-0.191]   &      &      &      &   [0.012]   &     \\
\noalign{\smallskip}Emotional IPV  & 0.433 & 0.527 & -0.094 & 263 & 0.323 & 0.306 & 0.018 & 1,161\\
 &   (0.497)   &   (0.501)   &   (0.061)   &      &   (0.468)   &   (0.461)   &   (0.027)   &     \\
 &      &      &   [-0.189]   &      &      &      &   [0.038]   &     \\
\noalign{\smallskip}Physical/sexual IPV  & 0.157 & 0.186 & -0.029 & 263 & 0.125 & 0.137 & -0.012 & 1,159\\
 &   (0.365)   &   (0.391)   &   (0.046)   &      &   (0.331)   &   (0.345)   &   (0.020)   &     \\
 &      &      &   [-0.078]   &      &      &      &   [-0.037]   &     \\
\noalign{\smallskip}\textbf{Joint significance (\textit{p}-value)} &  &  & 0.544 &  &  &  & 0.631 & \\
\noalign{\smallskip}\hline\end{tabular}\\

%% file: po_time_m_mon_cohab_ex1_2.tex
{
\def\sym#1{\ifmmode^{#1}\else\(^{#1}\)\fi}
\begin{tabular}{l*{6}{c}}
\toprule 
\noalign{\smallskip} 
                &\multicolumn{2}{c}{Any}  &\multicolumn{2}{c}{Emotional}&\multicolumn{2}{c}{Physical/Sexual}\\\cmidrule(lr){2-3}\cmidrule(lr){4-5}\cmidrule(lr){6-7}
                &\multicolumn{1}{c}{(1)}&\multicolumn{1}{c}{(2)}&\multicolumn{1}{c}{(3)}&\multicolumn{1}{c}{(4)}&\multicolumn{1}{c}{(5)}&\multicolumn{1}{c}{(6)}\\
                &\multicolumn{1}{c}{Near}&\multicolumn{1}{c}{Far}&\multicolumn{1}{c}{Near}&\multicolumn{1}{c}{Far}&\multicolumn{1}{c}{Near}&\multicolumn{1}{c}{Far}\\\noalign{\smallskip} 
 \midrule 
\noalign{\smallskip}
IPV Recall      &   -1.417   &   -5.241   &   -2.336   &   -5.527   &    2.751   &   -0.293   \\
                &  (4.264)   &  (4.304)   &  (4.172)   &  (4.197)   &  (3.663)   &  (3.596)   \\
\noalign{\medskip} 
IPV             &   -3.539   &   -8.937*  &   -4.378   &   -9.028*  &   -3.239   &   -7.535   \\
                &  (4.849)   &  (4.599)   &  (4.895)   &  (4.636)   &  (6.414)   &  (5.889)   \\
\noalign{\medskip} 
IPV Recall X IPV&    6.706   &    13.57** &    10.80   &    16.39** &   -12.72   &  -0.0837   \\
                &  (6.952)   &  (6.656)   &  (7.110)   &  (6.800)   &  (8.791)   &  (8.572)   \\
\noalign{\smallskip} 
\midrule
\noalign{\smallskip} 
Observations    &     1421   &     1421   &     1424   &     1424   &     1422   &     1422   \\
Control Mean    &    89.19   &    88.62   &    89.19   &    88.62   &    89.19   &    88.62   \\
Coef: Recall + Recall X IPV&    5.289   &    8.331   &    8.466   &   10.866   &   -9.971   &   -0.377   \\
p\_val: Recall + Recall X IPV&    0.334   &    0.099   &    0.140   &    0.041   &    0.213   &    0.961   \\
\noalign{\smallskip} 
\bottomrule
\end{tabular}
}

%% file: first_stage_vil_hrs_13122024.tex
{
\def\sym#1{\ifmmode^{#1}\else\(^{#1}\)\fi}
\begin{tabular}{l*{6}{c}}
\toprule
                &\multicolumn{2}{c}{Any}  &\multicolumn{2}{c}{Emotional}&\multicolumn{2}{c}{Physical/Sexual}\\\cmidrule(lr){2-3}\cmidrule(lr){4-5}\cmidrule(lr){6-7}
                &\multicolumn{1}{c}{(1)}&\multicolumn{1}{c}{(2)}&\multicolumn{1}{c}{(3)}&\multicolumn{1}{c}{(4)}&\multicolumn{1}{c}{(5)}&\multicolumn{1}{c}{(6)}\\
                &\multicolumn{1}{c}{Dummy}&\multicolumn{1}{c}{Index}&\multicolumn{1}{c}{Dummy}&\multicolumn{1}{c}{Index}&\multicolumn{1}{c}{Dummy}&\multicolumn{1}{c}{Index}\\
\midrule
Cash Transfer   &   -0.015   &   -0.582***&    0.009   &   -0.134   &   -0.077** &   -0.449***\\
                &  (0.013)   &  (0.206)   &  (0.022)   &  (0.086)   &  (0.033)   &  (0.156)   \\
\midrule
Observations    &     1010   &     1010   &     1010   &     1010   &     1010   &     1010   \\
Control Mean    &    0.965   &    3.099   &    0.897   &    1.692   &    0.378   &    1.407   \\
Effective F-stat&    1.357   &    8.016   &    0.180   &    2.404   &    5.629   &    8.303   \\
\bottomrule
\end{tabular}
}

%% file: 2sls_treat_vil_hrs_13122024_ar.tex
{
\def\sym#1{\ifmmode^{#1}\else\(^{#1}\)\fi}
\begin{tabular}{l*{6}{c}}
\toprule 
\noalign{\smallskip} 
 &\multicolumn{4}{c}{Index} &\multicolumn{2}{c}{Dummy}\\\cmidrule(lr){2-5}\cmidrule(lr){6-7}
                &\multicolumn{2}{c}{Any}&\multicolumn{2}{c}{Physical/Sexual}&\multicolumn{2}{c}{Physical/Sexual}\\\cmidrule(lr){2-3}\cmidrule(lr){4-5}\cmidrule(lr){6-7}
                &\multicolumn{1}{c}{(1)}&\multicolumn{1}{c}{(2)}&\multicolumn{1}{c}{(3)}&\multicolumn{1}{c}{(4)}&\multicolumn{1}{c}{(5)}&\multicolumn{1}{c}{(6)}\\
                &\multicolumn{1}{c}{Near}&\multicolumn{1}{c}{Far}&\multicolumn{1}{c}{Near}&\multicolumn{1}{c}{Far}&\multicolumn{1}{c}{Near}&\multicolumn{1}{c}{Far}\\
\noalign{\smallskip} 
 \midrule 
\noalign{\smallskip}
IPV reduction       &    8.955*  &   10.320   &   11.621*  &   13.393   &   67.403   &   77.679   \\
                &  (0.088)   &  (0.125)   &  (0.090)   &  (0.121)   &  (0.140)   &  (0.139)   \\
\noalign{\smallskip} 
\midrule
\noalign{\smallskip} 
Observations    &     1010   &     1010   &     1010   &     1010   &     1010   &     1010   \\
Control Mean    &   44.609   &   44.501   &   44.609   &   44.501   &   44.609   &   44.501   \\
Effective F-stat&    8.016   &    8.016   &    8.303   &    8.303   &    5.629   &    5.629   \\
Anderson-Rubin 90\% CI&   [ 1.46, 24.11]    &     [-0.20, 28.45]   &    [ 1.82, 31.45]    &    [-0.11, 36.10]   & [ 11.19,  $\infty$ ]   & [ -0.85,   $\infty$ ]   \\
\noalign{\smallskip} 
\bottomrule
\end{tabular}
}

%% file: first_stage_allarms.tex
{
\def\sym#1{\ifmmode^{#1}\else\(^{#1}\)\fi}
\begin{tabular}{l*{6}{c}}
\toprule 
\noalign{\smallskip} 
                &\multicolumn{2}{c}{Any}  &\multicolumn{2}{c}{Emotional}&\multicolumn{2}{c}{Physical/Sexual}\\\cmidrule(lr){2-3}\cmidrule(lr){4-5}\cmidrule(lr){6-7}
                &\multicolumn{1}{c}{(1)}&\multicolumn{1}{c}{(2)}&\multicolumn{1}{c}{(3)}&\multicolumn{1}{c}{(4)}&\multicolumn{1}{c}{(5)}&\multicolumn{1}{c}{(6)}\\
                &\multicolumn{1}{c}{Dummy}&\multicolumn{1}{c}{Index}&\multicolumn{1}{c}{Dummy}&\multicolumn{1}{c}{Index}&\multicolumn{1}{c}{Dummy}&\multicolumn{1}{c}{Index}\\
\midrule
Psychotherapy &  -0.0805** &   -0.159   &  -0.0649** &   -0.129   &  -0.0311   &  -0.0333   \\
                & (0.0319)   &  (0.128)   & (0.0294)   & (0.0794)   & (0.0198)   & (0.0595)   \\
\midrule
Observations    &     1158   &     1158   &     1161   &     1161   &     1159   &     1159   \\
Control Mean    &     0.38   &     1.06   &     0.35   &     0.73   &     0.15   &     0.33   \\
Effective F-stat&     6.39   &     1.56   &     4.88   &     2.62   &     2.48   &     0.31   \\
\bottomrule
\end{tabular}
}

%% file: 2sls_treat_vil_allarms.tex
{
\def\sym#1{\ifmmode^{#1}\else\(^{#1}\)\fi}
\begin{tabular}{l*{4}{c}}
\toprule 
\noalign{\smallskip} 
 &\multicolumn{4}{c}{Dummy}  \\\cmidrule(lr){2-5}
                &\multicolumn{2}{c}{Any}  &\multicolumn{2}{c}{Emotional}\\\cmidrule(lr){2-3}\cmidrule(lr){4-5}
                &\multicolumn{1}{c}{(1)}&\multicolumn{1}{c}{(2)}&\multicolumn{1}{c}{(3)}&\multicolumn{1}{c}{(4)}\\
                &\multicolumn{1}{c}{Near}&\multicolumn{1}{c}{Far}&\multicolumn{1}{c}{Near}&\multicolumn{1}{c}{Far}\\
\midrule
IPV reduction           &  -58.168   &  -76.827   &  -65.867   &  -88.869   \\
                &  (0.247)   &  (0.119)   &  (0.284)   &  (0.143)   \\
\midrule
Observations    &     1158   &     1158   &     1161   &     1161   \\
Control Mean    &   83.036   &   78.585   &   83.036   &   78.585   \\
Effective F-stat&    6.387   &    6.387   &    4.882   &    4.882   \\
\bottomrule
\end{tabular}
}

%% file: balance_ex1_normdiff.tex
\begin{tabular}{lccc}
\hline \noalign{\smallskip} & IPV Recall & Financial Concerns Recall & IPV Recall\\
 & versus & versus & versus\\
 & Placebo Recall & Placebo Recall & Financial Concerns Recall\\
\noalign{\smallskip}\hline \noalign{\smallskip}Age & 0.864 & 0.054 & 0.810\\
 & \begin{footnotesize}(0.300)\end{footnotesize} & \begin{footnotesize}(0.901)\end{footnotesize} & \begin{footnotesize}(0.901)\end{footnotesize}\\
 & \begin{footnotesize}[0.120]\end{footnotesize} & \begin{footnotesize}[0.008]\end{footnotesize} & \begin{footnotesize}[0.107]\end{footnotesize}\\
\noalign{\smallskip}Years of education & -0.101 & -0.583 & 0.482\\
 & \begin{footnotesize}(0.789)\end{footnotesize} & \begin{footnotesize}(0.152)\end{footnotesize} & \begin{footnotesize}(0.152)\end{footnotesize}\\
 & \begin{footnotesize}[-0.028]\end{footnotesize} & \begin{footnotesize}[-0.157]\end{footnotesize} & \begin{footnotesize}[0.131]\end{footnotesize}\\
\noalign{\smallskip}Children less than 6 & -0.107 & -0.193 & 0.086\\
 & \begin{footnotesize}(0.214)\end{footnotesize} & \begin{footnotesize}(0.030)**\end{footnotesize} & \begin{footnotesize}(0.030)**\end{footnotesize}\\
 & \begin{footnotesize}[-0.148]\end{footnotesize} & \begin{footnotesize}[-0.264]\end{footnotesize} & \begin{footnotesize}[0.119]\end{footnotesize}\\
\noalign{\smallskip}Children more than 6 & -0.078 & 0.007 & -0.085\\
 & \begin{footnotesize}(0.481)\end{footnotesize} & \begin{footnotesize}(0.965)\end{footnotesize} & \begin{footnotesize}(0.965)\end{footnotesize}\\
 & \begin{footnotesize}[-0.084]\end{footnotesize} & \begin{footnotesize}[0.007]\end{footnotesize} & \begin{footnotesize}[-0.084]\end{footnotesize}\\
\noalign{\smallskip}Years cohabitated & 0.307 & -0.406 & 0.713\\
 & \begin{footnotesize}(0.742)\end{footnotesize} & \begin{footnotesize}(0.664)\end{footnotesize} & \begin{footnotesize}(0.664)\end{footnotesize}\\
 & \begin{footnotesize}[0.040]\end{footnotesize} & \begin{footnotesize}[-0.057]\end{footnotesize} & \begin{footnotesize}[0.091]\end{footnotesize}\\
\noalign{\smallskip}Age (partner) & 0.436 & -1.116 & 1.552\\
 & \begin{footnotesize}(0.631)\end{footnotesize} & \begin{footnotesize}(0.279)\end{footnotesize} & \begin{footnotesize}(0.279)\end{footnotesize}\\
 & \begin{footnotesize}[0.049]\end{footnotesize} & \begin{footnotesize}[-0.141]\end{footnotesize} & \begin{footnotesize}[0.176]\end{footnotesize}\\
\noalign{\smallskip}Years of education (partner) & -0.343 & -0.785 & 0.442\\
 & \begin{footnotesize}(0.469)\end{footnotesize} & \begin{footnotesize}(0.137)\end{footnotesize} & \begin{footnotesize}(0.137)\end{footnotesize}\\
 & \begin{footnotesize}[-0.078]\end{footnotesize} & \begin{footnotesize}[-0.173]\end{footnotesize} & \begin{footnotesize}[0.096]\end{footnotesize}\\
\noalign{\smallskip}Married & 0.033 & 0.020 & 0.013\\
 & \begin{footnotesize}(0.298)\end{footnotesize} & \begin{footnotesize}(0.556)\end{footnotesize} & \begin{footnotesize}(0.556)\end{footnotesize}\\
 & \begin{footnotesize}[0.130]\end{footnotesize} & \begin{footnotesize}[0.074]\end{footnotesize} & \begin{footnotesize}[0.057]\end{footnotesize}\\
\noalign{\smallskip}Migrant & -0.039 & 0.014 & -0.054\\
 & \begin{footnotesize}(0.483)\end{footnotesize} & \begin{footnotesize}(0.735)\end{footnotesize} & \begin{footnotesize}(0.735)\end{footnotesize}\\
 & \begin{footnotesize}[-0.093]\end{footnotesize} & \begin{footnotesize}[0.032]\end{footnotesize} & \begin{footnotesize}[-0.125]\end{footnotesize}\\
\noalign{\smallskip}No formal job (partner) & -0.037 & -0.042 & 0.005\\
 & \begin{footnotesize}(0.607)\end{footnotesize} & \begin{footnotesize}(0.536)\end{footnotesize} & \begin{footnotesize}(0.536)\end{footnotesize}\\
 & \begin{footnotesize}[-0.075]\end{footnotesize} & \begin{footnotesize}[-0.085]\end{footnotesize} & \begin{footnotesize}[0.010]\end{footnotesize}\\
\noalign{\smallskip}Any IPV & -0.095 & 0.026 & -0.121\\
 & \begin{footnotesize}(0.114)\end{footnotesize} & \begin{footnotesize}(0.686)\end{footnotesize} & \begin{footnotesize}(0.686)\end{footnotesize}\\
 & \begin{footnotesize}[-0.191]\end{footnotesize} & \begin{footnotesize}[0.052]\end{footnotesize} & \begin{footnotesize}[-0.244]\end{footnotesize}\\
\noalign{\smallskip}Emotional IPV & -0.094 & 0.050 & -0.144\\
 & \begin{footnotesize}(0.125)\end{footnotesize} & \begin{footnotesize}(0.423)\end{footnotesize} & \begin{footnotesize}(0.423)\end{footnotesize}\\
 & \begin{footnotesize}[-0.189]\end{footnotesize} & \begin{footnotesize}[0.099]\end{footnotesize} & \begin{footnotesize}[-0.290]\end{footnotesize}\\
\noalign{\smallskip}Physical/sexual IPV & -0.029 & 0.011 & -0.040\\
 & \begin{footnotesize}(0.521)\end{footnotesize} & \begin{footnotesize}(0.816)\end{footnotesize} & \begin{footnotesize}(0.816)\end{footnotesize}\\
 & \begin{footnotesize}[-0.078]\end{footnotesize} & \begin{footnotesize}[0.028]\end{footnotesize} & \begin{footnotesize}[-0.106]\end{footnotesize}\\
\noalign{\smallskip}\textbf{Joint significance (p-value)} & 0.558 & 0.154 & 0.229\\
\noalign{\smallskip}\hline\end{tabular}\\

%% file: po_time_m_mon_cohab_mswitch_ex1_2.tex
{
\def\sym#1{\ifmmode^{#1}\else\(^{#1}\)\fi}
\begin{tabular}{l*{6}{c}}
\toprule 
\noalign{\smallskip} 
                &\multicolumn{2}{c}{Any}  &\multicolumn{2}{c}{Emotional}&\multicolumn{2}{c}{Physical/Sexual}\\\cmidrule(lr){2-3}\cmidrule(lr){4-5}\cmidrule(lr){6-7}
                &\multicolumn{1}{c}{(1)}&\multicolumn{1}{c}{(2)}&\multicolumn{1}{c}{(3)}&\multicolumn{1}{c}{(4)}&\multicolumn{1}{c}{(5)}&\multicolumn{1}{c}{(6)}\\
                &\multicolumn{1}{c}{Near}&\multicolumn{1}{c}{Far}&\multicolumn{1}{c}{Near}&\multicolumn{1}{c}{Far}&\multicolumn{1}{c}{Near}&\multicolumn{1}{c}{Far}\\
\noalign{\smallskip} 
 \midrule 
\noalign{\smallskip}
IPV Recall      &   -1.465   &   -5.319   &   -2.389   &   -5.602   &    2.714   &   -0.345   \\
                &  (4.288)   &  (4.309)   &  (4.195)   &  (4.201)   &  (3.684)   &  (3.604)   \\
\noalign{\medskip} 
IPV             &   -3.592   &   -8.947*  &   -4.406   &   -9.032*  &   -3.482   &   -7.611   \\
                &  (4.872)   &  (4.614)   &  (4.919)   &  (4.651)   &  (6.423)   &  (5.890)   \\

\noalign{\medskip} 
IPV Recall X IPV&    6.711   &    13.68** &    10.83   &    16.51** &   -12.71   &  -0.0596   \\
                &  (6.979)   &  (6.673)   &  (7.138)   &  (6.818)   &  (8.808)   &  (8.574)   \\
\noalign{\smallskip} 
\midrule
\noalign{\smallskip} 
Observations    &     1413   &     1418   &     1416   &     1421   &     1414   &     1419   \\
Control Mean    &    89.19   &    88.62   &    89.19   &    88.62   &    89.19   &    88.62   \\
Coef: Recall + Recall X IPV&    5.247   &    8.361   &    8.441   &   10.907   &   -9.996   &   -0.405   \\
p\_val: Recall + Recall X IPV&    0.340   &    0.099   &    0.142   &    0.041   &    0.212   &    0.958   \\
\noalign{\smallskip} 
\bottomrule
\end{tabular}
}

%% file: po_time_m_mon_discount_cohab_ex1_2.tex
{
\def\sym#1{\ifmmode^{#1}\else\(^{#1}\)\fi}
\begin{tabular}{l*{6}{c}}
\toprule 
\noalign{\smallskip} 
                &\multicolumn{2}{c}{Any}  &\multicolumn{2}{c}{Emotional}&\multicolumn{2}{c}{Physical/Sexual}\\\cmidrule(lr){2-3}\cmidrule(lr){4-5}\cmidrule(lr){6-7}
                &\multicolumn{1}{c}{(1)}&\multicolumn{1}{c}{(2)}&\multicolumn{1}{c}{(3)}&\multicolumn{1}{c}{(4)}&\multicolumn{1}{c}{(5)}&\multicolumn{1}{c}{(6)}\\
                &\multicolumn{1}{c}{Near}&\multicolumn{1}{c}{Far}&\multicolumn{1}{c}{Near}&\multicolumn{1}{c}{Far}&\multicolumn{1}{c}{Near}&\multicolumn{1}{c}{Far}\\
\noalign{\smallskip} 
 \midrule 
\noalign{\smallskip}
IPV Recall      &  -0.0021   &  -0.0058   &  -0.0029   &  -0.0058   &   0.0015   &  -0.0016   \\
                & (0.0047)   & (0.0047)   & (0.0046)   & (0.0046)   & (0.0039)   & (0.0038)   \\
\noalign{\medskip} 
IPV             &  -0.0032   &  -0.0079*  &  -0.0039   &  -0.0081*  &  -0.0036   &  -0.0087   \\
                & (0.0050)   & (0.0047)   & (0.0050)   & (0.0047)   & (0.0066)   & (0.0059)   \\
\noalign{\medskip} 
IPV Recall X IPV&   0.0059   &    0.012*  &   0.0096   &    0.015** &   -0.011   &   0.0024   \\
                & (0.0073)   & (0.0069)   & (0.0074)   & (0.0070)   & (0.0093)   & (0.0087)   \\
\noalign{\smallskip} 
\midrule
\noalign{\smallskip}
Observations    &     1421   &     1421   &     1424   &     1424   &     1422   &     1422   \\
Control Mean    &     1.06   &     1.06   &     1.06   &     1.06   &     1.06   &     1.06   \\
Coef: Recall + Recall X IPV&    0.004   &    0.007   &    0.007   &    0.009   &   -0.009   &    0.001   \\
p\_val: Recall + Recall X IPV&    0.495   &    0.181   &    0.245   &    0.091   &    0.276   &    0.911   \\
\noalign{\smallskip} 
\bottomrule
\end{tabular}
}

%% file: po_time_m_mon_ever_cohab_ex1_2.tex
{
\def\sym#1{\ifmmode^{#1}\else\(^{#1}\)\fi}
\begin{tabular}{l*{6}{c}}
\toprule 
\noalign{\smallskip} 
                &\multicolumn{2}{c}{Any}  &\multicolumn{2}{c}{Emotional}&\multicolumn{2}{c}{Physical/Sexual}\\\cmidrule(lr){2-3}\cmidrule(lr){4-5}\cmidrule(lr){6-7}
                &\multicolumn{1}{c}{(1)}&\multicolumn{1}{c}{(2)}&\multicolumn{1}{c}{(3)}&\multicolumn{1}{c}{(4)}&\multicolumn{1}{c}{(5)}&\multicolumn{1}{c}{(6)}\\
                &\multicolumn{1}{c}{Near}&\multicolumn{1}{c}{Far}&\multicolumn{1}{c}{Near}&\multicolumn{1}{c}{Far}&\multicolumn{1}{c}{Near}&\multicolumn{1}{c}{Far}\\\noalign{\smallskip} 
 \midrule 
\noalign{\smallskip}
IPV Recall      &   -0.345   &   -2.660   &   -0.432   &   -2.510   &    1.869   &   -0.833   \\
                &  (4.185)   &  (4.173)   &  (4.020)   &  (3.992)   &  (3.477)   &  (3.416)   \\
\noalign{\medskip} 
IPV             &   -4.439   &   -6.800   &   -3.467   &   -5.071   &   -6.813   &   -11.11** \\
                &  (4.320)   &  (4.175)   &  (4.370)   &  (4.216)   &  (4.968)   &  (4.606)   \\
\noalign{\medskip} 
IPV Recall X IPV&    4.371   &    6.025   &    5.688   &    6.943   &   -1.285   &    4.173   \\
                &  (6.110)   &  (5.894)   &  (6.184)   &  (5.968)   &  (6.976)   &  (6.610)   \\
\noalign{\smallskip} 
\midrule
\noalign{\smallskip} 
Observations    &     1736   &     1736   &     1740   &     1740   &     1737   &     1737   \\
Control Mean    &    89.19   &    88.62   &    89.19   &    88.62   &    89.19   &    88.62   \\
Coef: Recall + Recall X IPV&    4.027   &    3.365   &    5.256   &    4.433   &    0.584   &    3.340   \\
p\_val: Recall + Recall X IPV&    0.364   &    0.416   &    0.262   &    0.315   &    0.923   &    0.553   \\
\noalign{\smallskip} 
\bottomrule
\end{tabular}
}

%% file: po_risk_cf_miss_cohab_ex1_2.tex
{
\def\sym#1{\ifmmode^{#1}\else\(^{#1}\)\fi}
\begin{tabular}{l*{6}{c}}
\toprule 
\noalign{\smallskip} 
                &\multicolumn{2}{c}{Any}  &\multicolumn{2}{c}{Emotional}&\multicolumn{2}{c}{Physical/Sexual}\\\cmidrule(lr){2-3}\cmidrule(lr){4-5}\cmidrule(lr){6-7}
                &\multicolumn{1}{c}{(1)}&\multicolumn{1}{c}{(2)}&\multicolumn{1}{c}{(3)}&\multicolumn{1}{c}{(4)}&\multicolumn{1}{c}{(5)}&\multicolumn{1}{c}{(6)}\\
                &\multicolumn{1}{c}{Risk}&\multicolumn{1}{c}{Digit Span}&\multicolumn{1}{c}{Risk}&\multicolumn{1}{c}{Digit Span}&\multicolumn{1}{c}{Risk}&\multicolumn{1}{c}{Digit Span}\\
\noalign{\smallskip} 
 \midrule 
\noalign{\smallskip}
IPV-Recall      &    2.759   &  -0.0452   &    2.762   &  -0.0393   &    2.380   &  -0.0219   \\
                &  (1.905)   & (0.0656)   &  (1.871)   & (0.0645)   &  (1.594)   & (0.0555)   \\
\noalign{\medskip} 
IPV             &    1.235   &  -0.0670   &    1.272   &  -0.0682   &   -1.378   &   -0.153   \\
                &  (2.119)   & (0.0814)   &  (2.118)   & (0.0828)   &  (2.890)   &  (0.117)   \\
\noalign{\medskip} 
IPV-Recall X IPV&   -0.340   &    0.127   &   -0.844   &    0.120   &    1.657   &    0.163   \\
                &  (2.982)   &  (0.108)   &  (3.003)   &  (0.110)   &  (4.113)   &  (0.160)   \\
Strata          &      Yes   &      Yes   &      Yes   &      Yes   &      Yes   &      Yes   \\
\noalign{\smallskip} 
\midrule
\noalign{\smallskip} 
Observations    &     1421   &     1421   &     1424   &     1424   &     1422   &     1422   \\
Control Mean    &    47.16   &     2.61   &    47.16   &     2.61   &    47.16   &     2.61   \\
Coef: Recall + Recall X IPV&    2.419   &    0.082   &    1.918   &    0.081   &    4.037   &    0.142   \\
p-val: Recall + Recall X IPV&    0.289   &    0.340   &    0.412   &    0.363   &    0.284   &    0.346   \\
\noalign{\smallskip} 
\bottomrule
\end{tabular}
}

%% file: po_time_m_mon_cohab_ex1_before0.tex
{
\def\sym#1{\ifmmode^{#1}\else\(^{#1}\)\fi}
\begin{tabular}{l*{4}{c}}
\toprule 
\noalign{\smallskip} 
                &\multicolumn{2}{c}{Any}  &\multicolumn{2}{c}{Emotional}\\\cmidrule(lr){2-3}\cmidrule(lr){4-5}
                &\multicolumn{1}{c}{(1)}&\multicolumn{1}{c}{(2)}&\multicolumn{1}{c}{(3)}&\multicolumn{1}{c}{(4)}\\
                &\multicolumn{1}{c}{Near}&\multicolumn{1}{c}{Far}&\multicolumn{1}{c}{Near}&\multicolumn{1}{c}{Far}\\
\noalign{\smallskip} 
 \midrule 
\noalign{\smallskip}
IPV-Recall      &    20.62   &   -1.261   &    18.51   &   -1.603   \\
                &  (19.95)   &  (20.47)   &  (19.22)   &  (19.61)   \\
\noalign{\medskip} 
IPV             &   -1.566   &   -11.15   &   -6.014   &   -12.88   \\
                &  (19.03)   &  (19.95)   &  (18.14)   &  (18.94)   \\
\noalign{\medskip} 
IPV-Recall X IPV&    12.25   &    38.09   &    16.76   &    40.13   \\
                &  (26.52)   &  (26.69)   &  (26.04)   &  (26.14)   \\
\noalign{\smallskip} 
\midrule
\noalign{\smallskip} 
Observations    &      139   &      139   &      139   &      139   \\
Control Mean    &   108.12   &   122.17   &   108.12   &   122.17   \\
Coef: Recall + Recall X IPV&   32.877   &   36.832   &   35.268   &   38.525   \\
p-val: Recall + Recall X IPV&    0.070   &    0.037   &    0.054   &    0.031   \\
\noalign{\smallskip} 
\bottomrule
\end{tabular}
}

%% file: po_time_m_mon_cohab_ex2_cash_only.tex
{
\def\sym#1{\ifmmode^{#1}\else\(^{#1}\)\fi}
\begin{tabular}{l*{4}{c}}
\toprule 
\noalign{\smallskip} 
                &\multicolumn{2}{c}{Any}  &\multicolumn{2}{c}{Emotional}\\\cmidrule(lr){2-3}\cmidrule(lr){4-5}
                &\multicolumn{1}{c}{(1)}&\multicolumn{1}{c}{(2)}&\multicolumn{1}{c}{(3)}&\multicolumn{1}{c}{(4)}\\
                &\multicolumn{1}{c}{Near}&\multicolumn{1}{c}{Far}&\multicolumn{1}{c}{Near}&\multicolumn{1}{c}{Far}\\
\noalign{\smallskip} 
 \midrule 
\noalign{\smallskip}
IPV-Recall      &   -0.836   &   -7.360   &   -1.870   &   -7.212   \\
                &  (6.886)   &  (6.910)   &  (6.687)   &  (6.665)   \\
\noalign{\medskip} 
IPV             &   -6.702   &   -15.95** &   -7.098   &   -14.56** \\
                &  (7.653)   &  (6.891)   &  (7.767)   &  (7.066)   \\
\noalign{\medskip} 
IPV-Recall X IPV&    4.456   &    17.47*  &    8.152   &    18.50*  \\
                &  (11.24)   &  (10.41)   &  (11.59)   &  (10.77)   \\
\noalign{\smallskip} 
\midrule
\noalign{\smallskip} 
Observations    &      581   &      581   &      581   &      581   \\
Control Mean    &    86.90   &    83.83   &    86.90   &    83.83   \\
Coef: Recall + Recall X IPV&    3.620   &   10.111   &    6.282   &   11.287   \\
p-val: Recall + Recall X IPV&    0.680   &    0.184   &    0.500   &    0.172   \\
\noalign{\smallskip} 
\bottomrule
\end{tabular}
}

%% file: liml_treat_vil_hrs_13122024.tex
{
\def\sym#1{\ifmmode^{#1}\else\(^{#1}\)\fi}
\begin{tabular}{l*{6}{c}}
\toprule 
\noalign{\smallskip} 
            &\multicolumn{4}{c}{Index} &\multicolumn{2}{c}{Dummy}\\\cmidrule(lr){2-5}\cmidrule(lr){6-7}
                &\multicolumn{2}{c}{Any}&\multicolumn{2}{c}{Physical/Sexual}&\multicolumn{2}{c}{Physical/Sexual}\\\cmidrule(lr){2-3}\cmidrule(lr){4-5}\cmidrule(lr){6-7}
                &\multicolumn{1}{c}{(1)}&\multicolumn{1}{c}{(2)}&\multicolumn{1}{c}{(3)}&\multicolumn{1}{c}{(4)}&\multicolumn{1}{c}{(5)}&\multicolumn{1}{c}{(6)}\\
                &\multicolumn{1}{c}{Near}&\multicolumn{1}{c}{Far}&\multicolumn{1}{c}{Near}&\multicolumn{1}{c}{Far}&\multicolumn{1}{c}{Near}&\multicolumn{1}{c}{Far}\\
\noalign{\smallskip} 
\midrule
\noalign{\smallskip} 
IPV        &    8.955*  &   10.320   &   11.621*  &   13.393   &   67.403   &   77.679   \\
                &  (0.088)   &  (0.125)   &  (0.090)   &  (0.121)   &  (0.140)   &  (0.139)   \\
\noalign{\smallskip} 
\midrule
\noalign{\smallskip} 
Observations    &     1010   &     1010   &     1010   &     1010   &     1010   &     1010   \\
Control Mean    &   44.609   &   44.501   &   44.609   &   44.501   &   44.609   &   44.501   \\
Effective F-stat&    8.016   &    8.016   &    8.303   &    8.303   &    5.629   &    5.629   \\
\noalign{\smallskip} 
\bottomrule
\end{tabular}
}

%% file: exclusion_directtime_noIPVBL_GPM_allarms.tex
{
\def\sym#1{\ifmmode^{#1}\else\(^{#1}\)\fi}
\begin{tabular}{l*{2}{c}}
\toprule
                &\multicolumn{1}{c}{(1)}&\multicolumn{1}{c}{(2)}\\
                &\multicolumn{1}{c}{Near}&\multicolumn{1}{c}{Far}\\
\midrule
Psychotherapy   &    3.020   &    4.801   \\
                &  (0.463)   &  (0.199)   \\
\midrule
Observations    &      893   &      893   \\
Control Mean    &    83.04   &    78.59   \\
\bottomrule
\end{tabular}
}

%% file: exclusion_stress_GPM_allarms.tex
{
\def\sym#1{\ifmmode^{#1}\else\(^{#1}\)\fi}
\begin{tabular}{l*{1}{c}}
\toprule
                &\multicolumn{1}{c}{(1)}\\
                &\multicolumn{1}{c}{Stress}\\
\midrule
Psychotherapy   &    0.038   \\
                &  (0.626)   \\
\midrule
Observations    &     1111   \\
Control Mean    &    0.000   \\
\bottomrule
\end{tabular}
}

%% file: exclusion_stress_HRS.tex
{
\def\sym#1{\ifmmode^{#1}\else\(^{#1}\)\fi}
\begin{tabular}{l*{1}{c}}
\toprule
                &\multicolumn{1}{c}{(1)}\\
                &\multicolumn{1}{c}{Stress}\\
\midrule
Cash Transfer   &   -0.155*  \\
                & (0.0581)   \\
\midrule
Observations    &     1189   \\
Control Mean    &    0.00   \\
\bottomrule
\end{tabular}
}

%% file: 2sls_stress_HRS.tex
{
\def\sym#1{\ifmmode^{#1}\else\(^{#1}\)\fi}
\begin{tabular}{l*{2}{c}}
\toprule 
\noalign{\smallskip} 
                &\multicolumn{1}{c}{(1)}&\multicolumn{1}{c}{(2)}\\
                &\multicolumn{1}{c}{Near}&\multicolumn{1}{c}{Far}\\
\noalign{\smallskip} 
\midrule
\noalign{\smallskip} 
Stress reduction       &    6.469   &   10.019   \\
                &  (0.382)   &  (0.329)   \\
\noalign{\smallskip} 
\midrule
\noalign{\smallskip} 
Observations    &     1189   &     1189   \\
Control Mean    &   44.609   &   44.501   \\
Effective F-stat&    3.661   &    3.661   \\
\noalign{\smallskip} 
\bottomrule
\end{tabular}
}